\documentclass[10pt]{article}

\usepackage{graphics}

\newcommand{\mycomment}[1]{}
\usepackage{subcaption}

\usepackage{adjustbox}
\usepackage{multirow}

\usepackage{gensymb} 
\usepackage{siunitx} 
\usepackage{verbatim} 
\newcommand\Tstrut{\rule{0pt}{2.6ex}}         
\newcommand\Bstrut{\rule[-1.3ex]{0pt}{0pt}}   
\usepackage{standalone} 

\usepackage[section]{placeins}

\usepackage{tikz}
\usepackage{adjustbox}

\usepackage{mathtools}

\usepackage[round]{natbib}

\usepackage{amsmath, amsfonts, amssymb, amsthm, dsfont, mathtools}

\usepackage{mathrsfs}

\usepackage{soul}

\usepackage{bm}

\usepackage{booktabs} 
 
\usepackage{geometry}
\definecolor{mypurple}{RGB}{0,0,0}
\definecolor{myblue}{RGB}{0,87,120}
\definecolor{aqua}{RGB}{0,0,0}
\definecolor{myorange}{RGB}{0,0,0}
\definecolor{mygrey}{RGB}{255,255,255}

\usepackage{ulem}

\usepackage{graphicx}
\usepackage{color}
\usepackage{nicefrac}
\usepackage{booktabs}

\usepackage{titlesec}
\titleformat{\section}[hang]{\large\scshape}{\thesection.}{1em}{}
\titleformat{\subsection}[hang]{\large\scshape}{\thesubsection.}{1em}{}
\titleformat{\subsubsection}[hang]{\scshape}{\thesubsubsection.}{1em}{}

\usepackage[colorlinks=true, linkcolor=red, urlcolor=blue, citecolor=blue]{hyperref}

\newtheoremstyle{mytheoremstyle} 
    {0.3cm}                      
    {0cm}                        
    {\itshape}                   
    {}                           
    {\scshape}                   
    {: }                          
    {0em}                       
    {}  

\theoremstyle{mytheoremstyle}

\newtheorem{lemma}{Lemma}

\newtheoremstyle{mytheoremstyle} 
    {0.3cm}                      
    {0cm}                        
    {\itshape}                   
    {}                           
    {\scshape}                   
    {: }                          
    {0em}                       
    {}  

\theoremstyle{mytheoremstyle}
\newtheorem{Theorem}{Theorem}

\newtheoremstyle{myExampleRemarkstyle} 
    {0.3cm}                    
    {0cm}                           
    {\itshape}                   
    {}                           
    {\scshape}                   
    {: }                          
    {0em}                       
    {}  

\theoremstyle{myExampleRemarkstyle}
 
\newtheorem{Remark}{Remark}
\newtheorem{Assumption}{Assumption}

\newtheoremstyle{simuStyle}
{0.3cm} 
{0cm} 
{} 
{} 
{\bfseries} 
{.} 
{0em} 
{} 

\theoremstyle{simuStyle}

\newtheoremstyle{stratStyle}
{0.3cm} 
{0cm} 
{} 
{} 
{\scshape} 
{: } 
{0em} 
{} 

\theoremstyle{stratStyle}

\DeclareSymbolFont{lettersA}{U}{txmia}{m}{it}
\DeclareMathSymbol{\real}{\mathord}{lettersA}{"92}
\DeclareMathSymbol{\complex}{\mathord}{lettersA}{"83}

\def\real{{\rm I\!R}}

\def\0{{\bf 0}}

\def\btheta{{\bm{\theta}}}

\mathchardef\Re="023C
\mathchardef\Im="023D

\def\bt{\bm{\theta}}

\def\bT{\bm{\Theta}}

\def\0{\mathbf{0}}

\def\N{\mathbb{N}}
\def\Ns{\N^\ast}

\def\boxit#1{\vbox{\hrule\hbox{\vrule\kern3pt
          \vbox{\kern3pt#1\kern3pt}\kern3pt\vrule}\hrule}}
          
\def\lvcomment#1{\vskip0mm\boxit{\vskip 0mm{\color{violet}\bf#1}{\color{violet}\bf -- L.V.\vskip 1mm}}\vskip 0mm}

\def\mkcomment#1{\vskip0mm\boxit{\vskip 0mm{\color{violet}\bf#1}{\color{violet}\bf -- MK\vskip 1mm}}\vskip 0mm}

\definecolor{pinegreen}{rgb}{0.0, 0.47, 0.44}

\usepackage{setspace}	
\begin{document} 

\let\refBKP\ref
\renewcommand{\ref}[1]{{\upshape\refBKP{#1}}}

\begin{center}
\begin{doublespacing}
    {\huge \textsc{Accounting for Vibration Noise\\ in Stochastic Measurement Errors}}
	\\
\end{doublespacing}
 \vspace{0.75cm}
	{\textsc{Lionel Voirol}$^{\dag}$, \textsc{Davide A. Cucci}$^{\dag}$, 
    \textsc{Mucyo Karemera}$^{\dag}$,\\
    \textsc{Wenfei Chu}$^{\dag}$, \textsc{Roberto Molinari}$^{\ddag}$,   \& \textsc{St\'ephane Guerrier}}$^{\dag \, *}$\\
	\vspace{0.25cm}
	{$^{\dag}$Geneva School of Economics and Management, $^{*}$Faculty of Science, University of Geneva\\
	$^{\ddag}$ Department of Mathematics and Statistics, Auburn University}
	\vspace{0.5cm}
\end{center}

\textsc{Abstract:} The measurement of data over time and/or space is of utmost importance in a wide range of domains from engineering to physics. Devices that perform these measurements therefore need to be extremely precise to obtain correct system diagnostics and accurate predictions, consequently requiring a rigorous calibration procedure which models their errors before being employed. While the deterministic components of these errors do not represent a major modelling challenge, most of the research over the past years has focused on delivering methods that can explain and estimate the complex stochastic components of these errors. This effort has allowed to greatly improve the precision and uncertainty quantification of measurement devices but has this far not accounted for a significant stochastic noise that arises for many of these devices: vibration noise. Indeed, having filtered out physical explanations for this noise, a residual stochastic component often carries over which can drastically affect measurement precision. This component can originate from different sources, including the internal mechanics of the measurement devices as well as the movement of these devices when placed on moving objects or vehicles. To remove this disturbance from signals, this work puts forward a modelling framework for this specific type of noise and adapts the Generalized Method of Wavelet Moments to estimate these models. We deliver the asymptotic properties of this method when applied to processes that include vibration noise and show the considerable practical advantages of this approach in simulation and applied case studies.

\noindent \textsc{Keywords:} {Wavelet Variance, Inertial Measurement Unit, Generalized Method of Wavelet Moments, Stochastic Error}

\noindent \textsc{Acknowledgements}: S. Guerrier, D. A. Cucci, M. Karemera and W. Chu were supported by the SNSF Grants \#176843 and \#211007 and the Innosuisse Grants \#37308.1 IP-ENG and \#53622.1 IP-ENG. L. Voirol was supported by the SNSF Grant \#182684. Roberto Molinari was also partially supported by NSF Grant SES-2150615.

\section{Introduction}

The task of measuring and predicting the evolution of different physical systems passes through the precision of the instruments built to carry out such a task. In order to measure the evolution of these systems, the devices need to perform repeated measurements (often at high frequency) and can suffer from errors that can accumulate over time and, consequently, have extreme negative impacts in many fields \citep[see][]{titterton2004strapdown, webster2018measurement}. For this purpose, especially when dealing with high-precision measurement, these devices need to go through a rigorous calibration procedure which, in the majority of cases, is performed within a certain time-frame and in controlled (static) settings. These procedures are mainly aimed at quantifying and characterizing the measurement error of these devices so as to explain and model them, subsequently allowing to remove or filter out these errors when actually employed in the real world as well as to deliver reliable uncertainty metrics. An example of such a procedure can be found in inertial sensor calibration where these devices are widely and increasingly being employed in different areas, from robotics to unmanned navigation, because of their low cost and light weight \citep[see e.g.,][]{el2020inertial}. Due to these characteristics, inertial sensors often suffer from important measurement errors which, like many phenomena measured over time, have deterministic and stochastic components where the latter often have a considerable impact in the overall measurement error \citep[see e.g.,][]{el2007analysis}. Indeed, while various statistical or machine learning techniques can be employed to explain and remove the deterministic component based also on physical models, the stochastic component (which is the focus of this work) still represents a modelling challenge under many aspects and is essential to quantify the measurement uncertainty. 

More specifically, the stochastic measurement error of these instruments (such as that of inertial sensors) is frequently characterized by a complex spectral structure generally explained by \textit{composite} models that are constituted by the sum of different stochastic error processes which contribute to the overall observed error \citep[many state-space models take on this form, see e.g.,][]{el2007analysis, stebler2014generalized}. The different underlying (latent) stochastic error processes can either have a direct physical justification (such as a random error accumulation represented by a random walk) and/or can be extremely useful in closely approximating the overall error structure. An example of such composite processes is the renown class of Auto-Regressive-Moving-Average models which can be represented by the sum of individual white noise and first-order autoregressive processes, as well as the larger class of (linear) state-space models. It must be underlined that the stochastic errors of these devices are commonly observed in static conditions and, hence, the above composite models are chosen and estimated for this specific setting. However, in many applied settings, there are circumstances where there are additional noise components that are unaccounted for by these composite models. Indeed, there are often additional structures in the noise that, as a result of sources of vibration, are of cyclical nature and can arise for different reasons \citep[see e.g.,][]{1706054}. For example, the measurement devices themselves can produce vibrations due to their electric and mechanical characteristics when their power is on, thereby requiring the inclusion of an additional source of \textit{internal} noise to the composite model structure. Moreover, it has been highlighted how the stochastic properties of the device errors vary as a function of \textit{external} conditions, such as, for example, the ambient temperature and the sensor motion \citep[see e.g.,][]{stebler2015approach}. For instance, a device may exhibit higher noise levels or bias instability during highly dynamic motion or in presence of intense vibrations, see for example~\cite{radi2018stochastic}. More specifically, the impact of these dynamics on the measurement errors has been assessed in controlled environments through calibration instruments such as rotation or linear tables which are used to move these devices according to known and repeated patterns for a sufficient amount of time. Imperfections on the calibration instruments, such as rotation table control loops, make this process very difficult and spurious, entailing periodic disturbances that are left in the error signal and need to be removed through stochastic modeling.

To date, the estimation of the stochastic error component has been addressed by employing complex composite models which however do not include processes that describe the impact of vibrations on the measurements. In particular, even estimating these commonly employed composite models (without vibration) has represented an important computational challenge given the high-frequency and consequent length of the error signals that these devices record. For example, the Maximum Likelihood Estimator (MLE) is generally implemented through the use of an Extended Kalman Filter (EKF) and the Expectation-Maximization algorithm which both become numerically unstable or computationally prohibitive when considering the complexity of the composite models and the length of the error signals \citep[see][]{stebler2014generalized}. Other more tractable techniques have been proposed and adopted over the past years but they often lack adequate statistical properties \citep[see][]{guerrier2016theoretical}. For this purpose, \cite{guerrier2013wavelet} put forward the Generalized Method of Wavelet Moments (GMWM) which delivers a computationally feasible solution in these settings while preserving appropriate statistical properties. However, as stated earlier, none of these existing techniques have comprehensively addressed the presence of vibration noise in the stochastic signals. A substantial amount of literature has either underlined the need to find solutions to this problem or have put forward approaches that nevertheless do not adequately respond to the need to filter out this type of noise from device measurement errors. For example, in the context of stochastic calibration of Inertial Measurement Units (IMU), \cite{capriglione2019experimental} highlight that MicroElectroMechanical Systems (MEMS) are very sensitive to vibration via multiple experimental results and that adapted measurement procedures and corrections should be considered. Among the approaches which attempt to remove this source of noise,  \cite{gang2010detecting} propose a mathematical modelling of the vibration signal and discuss conditions under which it can be filtered out using wavelet transforms, while \cite{kang2013approach} propose a direct coning mitigation algorithm to test, estimate and compensate a sinusoidal component in the signal of a gyroscope whereas \cite{ma2017vibration} propose a gradient descent algorithm to correct vibration effects on an IMU. In all these cases the vibration error is treated as an ``outlier'' effect where the proposed methods aim to deliver adequate estimations that are in some way resistant to this effect. Hence, these methods are not tailored to the most common setting where the vibration noise is a structural component of the stochastic measurement errors in these devices. In this sense, to the best of the authors' knowledge, the only approach that aims at addressing this noise as a structural component of these measurement errors can be found in \cite{ng1997characterization} or \cite{Bos2008, bos2013fast} where however it is not considered as a stochastic component but as a deterministic one. Indeed, using the relationship between the Allan Variance (AV) and the spectral density function (see e.g., \citealp{el2007analysis}), in \cite{ng1997characterization} they derive a theoretical form of the AV for a sinusoidal process to consequently characterize the low- and high-frequency components of the measurement error in ring-laser gyroscopes. However this characterization was proven to be statistically inconsistent (see \citealp{guerrier2016theoretical}). Finally, like the MLE methodology in \cite{Bos2008,bos2013fast}, a limitation of these approaches is that the sinusoidal process is assumed to be deterministic and, as a consequence, certain characteristics of this process (such as its frequency or phase) are assumed to be known which is rarely the case in practice.


Following the above, in Section~\ref{sec:math_setup} this work firstly enriches the class of composite models used for stochastic modelling by adding one or more processes that adequately describe the impact of vibration on the stochastic error components of these measurement devices. Based on this, in Section~\ref{sec.method} it then aims to make use of the GMWM to estimate the composite models that include the vibration components, studying its properties in presence of vibration noise and giving theoretical guarantees that support the validity of this approach. Finally Section~\ref{section_simu} evaluates the numerical performance of the proposed method through different simulation studies while Section~\ref{section_application} highlights the advantages of this approach through the analysis of data issued from a low-cost inertial sensor in different dynamic conditions. Further discussions are presented in Section~\ref{section_conclusion}. The proofs of the theoretical results together with some additional information are collected in the appendix.



\section{Modeling Vibration Noise}\label{sec:math_setup}  

We first clearly define a model that can adequately describe such a stochastic vibration effect over time. For this purpose, let $(S_t), \,\, t \in \mathbb{Z}$, represent the process issued from the vibration source which we assume to be periodic. Based on this assumption, a natural candidate to model such a process is a wave function which, for this work, we parametrize as follows
\begin{equation} 
    S_{t} \vcentcolon= \alpha \sin (\beta t+U),
    \label{sin_process_def}
\end{equation}
with amplitude $\alpha\!\in\!\real_+\!\vcentcolon=\!(0,+\infty)$, angular frequency $\beta\in(0,\pi]$ and phase $U \sim \mathcal{U}(0, 2\pi)$. As confirmed by the application study presented in Section \ref{section_application} (as well as other applications not presented in this work), this parametrization of the vibration noise appears to well represent the true noise observed in real-life applications. Within this work, we will refer to the process as a \textit{sinusoidal} process. It is possible to notice that the only stochastic component in the above parametrization is the uniform random variable $U$ which randomly shifts the phase of the process. While other parametrizations of the trigonometric functions can obviously be considered, in many applications (including inertial measurement units) it is possible to observe roughly constant behaviours in terms of amplitude and frequency of the distinct vibration noises affecting each device, while the uncertainty usually lies in the phase of their periodicity.

Having defined the model to characterize vibration noise, we can increase its flexibility by considering a summation of $L\in\mathbb{N}$ independent latent sinusoidal processes given by 
\begin{equation} 
    \label{comp_sin_process_def}
    \sum_{l=1}^L S_{t}^{(l)} \vcentcolon= \sum_{l=1}^L \alpha_l \sin (\beta_l t + U_l),
\end{equation}
where the parameters and the random phase are now indexed by $l = 1, \hdots, L$ denoting how they specifically characterize the $l^{th}$ sinusoidal latent process. Indeed, due to the mechanical properties of a device or to the conditions under which it operates, there may be different sources of vibration affecting the measurement performance and such a composite process would address (or well approximate) the diversity of these vibration sources. 

As mentioned in the introduction, devices are characterized by a multitude of measurement errors which in many cases are modelled by composite processes which, like the one defined above, consist in a summation of different independent processes representing different sources of error. A general class of composite processes which is used to model these errors is given by:
\begin{equation*}
    Z_t \vcentcolon= W_t + Q_t + \sum_{k = 1}^K Y_{t}^{(k)} + D_t + R_t,
\end{equation*}
where $(W_t)$ represents a white noise (WN); $(Q_t)$ is a quantization noise (QN) which is a rounding error process (see \citealp{papoulis2002probability}); $(Y_t^{(k)})$ is the $k^{th}$ causal first-order autoregressive (AR1) process out of a total of $K\geq 1$ AR1 processes; $(D_t)$ is a deterministic drift (DR) process; and $(R_t)$ is a random walk (RW) process \citep[see e.g.,][for more details]{el2007analysis, guerrier2016theoretical}. Noting that the sum of AR1 and WN processes delivers an Auto-Regressive-Moving-Average (ARMA) process (see \citealp{granger1976time}), this class of composite processes is extremely flexible since the subsets of models that originate from it can adequately describe or approximate the behaviour of the vast majority stationary signals.

The goal of this work, as underlined in the introduction, is to reliably estimate the class of models $(Z_t)$ while accounting for vibration noise which would be considered as a ``nuisance'' process. As a consequence,  we aim to combine the above-defined classes of composite processes, i.e., $(S_t)$ and $(Z_t)$, to ensure that the additional sources of noise are addressed appropriately  by considering a larger class of processes which is intuitively given by their summation, i.e.,
\begin{equation}
    X_t \vcentcolon= W_t + Q_t + \sum_{k = 1}^K Y_{t}^{(k)} + D_t + R_t + \sum_{l=1}^L S_{t}^{(l)}.
    \label{eq:define:Xt}
\end{equation}
The first aspect to underline is that, from a practical perspective, one would not usually postulate a model based on \textit{all} the latent processes available in the class $(X_t)$. Nevertheless, in various applied settings it may be necessary to make use of at least one of each latent process since the noise characterizing measurement devices can have a highly complex spectral behaviour. In particular, for this work we consider the sum of vibration noises in \eqref{eq:define:Xt} to be a structural nuisance component that needs to be estimated to obtain adequate estimates of the processes that are specific to $(Z_t)$. In this optic, the next section studies how the proposed methodology estimates these nuisance components whose form is given in \eqref{sin_process_def}. 
%
%

\section{Estimation Framework}
\label{sec.method}


In this work we intend to make use of the GMWM which has been adapted to different time series estimation settings \citep[see e.g.,][]{xu2019multivariate, guerrier2020wavelet, guerrier2022robust}. To define this framework, let $F_{\btheta_0}$ represent the data-generating process with true parameter $\btheta_0 \in \bm{\Theta} \subset \real^p$ which we aim to estimate and perform inference on. The GMWM delivers an estimator of $\btheta_0$ based on the following generalized least-squares problem:
\begin{align}\label{def:GMWM}
      \hat{\bm{\theta}}\vcentcolon=\underset{\bm{\theta} \in \bm{\Theta}}{\operatorname{argmin}} \; \big\|\hat{\bm{\nu}}-\bm{\nu}(\bm{\theta})\big\|_{\bm{\Omega}}^{2} ,
\end{align}
where $\|\mathbf{x}\|_\mathbf{A}^2 \vcentcolon= \mathbf{x}^T\mathbf{A}\mathbf{x}$ with $\mathbf{x}\in\real^J$ and $\mathbf{A}\in\real^{J\times J}$; $\hat{\bm{\nu}} \in \real_+^J$ represents the Wavelet Variance (WV) estimated on the signal $(X_t)_{t=1,\dots, T}$; $\bm{\nu}(\bm{\theta}) = [\nu_j(\bm{\theta}]_{j=1,\dots J} \in \real_+^J$ is the theoretical WV implied by the model of interest; and $\bm{\Omega} \in \real^{J \times J}$ is a positive-definite weighting matrix for which a good choice, for example, is the inverse of the covariance matrix of $\hat{\bm{\nu}}$ \citep[see e.g.,][]{guerrier2013wavelet}. More specifically, the WV is the variance of the wavelet coefficients $(\omega_{j,t})$ issued from a wavelet decomposition of the process $(X_t)$ with relative scales of decomposition $j = 1. \hdots, J$, with $J < \log_2(T)$, and has different advantageous properties for the analysis of time series \citep[see e.g.,][]{serroukh2000statistical}.

The GMWM framework therefore relies firstly on the properties of the estimator of WV $\hat{\bm{\nu}}$, and then on those of the theoretical WV $\bm{\nu}(\bm{\theta})$ implied by the model $F_{\btheta}$. More specifically, the properties of the estimator $\hat{\bm{\nu}}$ were first studied in \cite{percival1995estimation} and \cite{serroukh2000statistical} under a set of standard conditions for time series analysis, followed by the results of \cite{xu2019multivariate} and \cite{guerrier2022robust} allowing for statistical consistency (and asymptotic normality) of this estimator under weaker conditions. For this reason, given the new process considered in this work, let us consider the properties of the estimator of WV $\hat{\bm{\nu}}$ when applied exclusively to the realization of a single sinusoidal process $(S_t)$ and denote this specific estimator as $\hat{\bm{\nu}}_S := [\hat{\nu}_{j,S}]_{j=1,\hdots,J}$ to indicate its implicit dependence on the parameters underlying this particular process. With $\tau_j$ denoting the length of the wavelet filter at the $j^{th}$ level of decomposition, the following lemma highlights the statistical consistency of $\hat{\bm{\nu}}_S$ when using the commonly used Haar wavelet filter for which $\tau_j = 2^j$.

\begin{lemma}\label{lemma:sin:asymp}
For the Haar wavelet filter and for any $j=1,\dots,J < \log_2(T)$, we have
\begin{equation*}
    \label{eq:theo:wv:sin}
    \hat{{\nu}}_{j,S} = {{\nu}}_{j}\left(\alpha, \beta\right) + \mathcal{O}_{p}\left(T^{-1}\right),
\end{equation*}
where
\begin{equation*}
\label{eq:theo:expect:wv:sin}
    {{\nu}}_{j}\left(\alpha, \beta\right) \vcentcolon= \mathbb{E}\left[\hat{{\nu}}_{j,S}\right] = \frac{\alpha^{2}\left\{1-\cos \left(\frac{\beta \tau_{j}}{2}\right)\right
\}^{2}}{\tau_{j}^{2}\{1-\cos \left(\beta\right)\}}.
\end{equation*}
\end{lemma}

The Haar wavelet filter is one of the most commonly employed wavelet filters \citep[see e.g.,][]{percival2000wavelet} and, as a result of this lemma (whose proof is given in Appendix \ref{appendix:proof:lemma:sin:asymp}), it is possible to see that the estimator $\hat{\bm{\nu}}_S$ is statistically consistent when making use of this filter. In addition, the use of the Haar filter is the only condition needed to obtain this result while also guaranteeing a convergence rate of $\mathcal{O}_p\left({T}^{-1}\right)$.


We now extend the study of the properties of the WV estimator $\hat{\bm{\nu}}$ to the class of models $(X_t)$ defined in \eqref{eq:define:Xt}, which adds the sinusoidal processes to the class of models $(Z_t)$. For this purpose we will first focus solely on defining the conditions required to obtain asymptotic normality of the estimator for the class of processes $(Z_t)$ and consequently denote the WV estimator applied exclusively to this class as $\hat{\bm{\nu}}_Z$. We will then combine these with the result of Lemma \ref{lemma:sin:asymp} to obtain the required properties for the class of processes of interest $(X_t)$. 

The properties of consistency and asymptotic normality of $\hat{\bm{\nu}}_Z$ have already been studied in \cite{xu2019multivariate} and \cite{guerrier2022robust} so, for the sake of completeness, we will briefly summarize and discuss the conditions needed to achieve these properties. To do so, let us start by denoting the first order difference of the process $(Z_t)$ as $\Delta_t \vcentcolon= Z_t -Z_{t-1}$. We also define $G(\cdot)$ to be an $\real$-valued measurable function as well the filtration $\mathcal{F}_t = (\dots,\epsilon_{t-1},\epsilon_t)$, where $\epsilon_t$ are i.i.d random variables. 
\begin{Assumption}\label{assum:delta:filtration}
The process $(\Delta_t)$ is strictly stationary and can be represented as $$\Delta_t = G(\mathcal{F}_t).$$
\end{Assumption}
This assumption is commonly required when analyzing time series and, in the setting of this work, allows to make use of the results in \cite{wu2011gaussian}. For the next assumptions, we also define the operation $\|D\|_{p} \vcentcolon=\left(\mathbb{E}[|D|^{p}]\right)^{1 / p}$, for $p>0$, as well as the filtration  $\mathcal{F}_t^\star = (\dots,\epsilon_0^\star,\dots,\epsilon_{t-1},\epsilon_t)$, where $\epsilon_0^\star$ is an i.i.d. random variable. The latter also allows us to define $\Delta_t^\star = G(\mathcal{F}_t^\star)$ which differs from $(\Delta_t)$ as a result of the different innovation noise at time $t=0$ (clearly we have $\Delta_t^\star = \Delta_t$ for $t<0$).

\begin{Assumption}\label{assum:moment:delta}
 $\left\|\Delta_{t}\right\|_{4}<\infty$. 
\end{Assumption}
%
%
\begin{Assumption}\label{assum:moment:delta:diff}
 $\sum_{t=0}^{\infty}\left\|\Delta_{t} - \Delta_{t}^{\star}\right\|_{4}<\infty$.
\end{Assumption}
%


To summarize, Assumptions \ref{assum:moment:delta} and \ref{assum:moment:delta:diff} require bounded fourth moments of the process $(\Delta_t)$ and of the difference between this process and its ``copy'', implying a stability of $(\Delta_t)$ since a change in the innovation process does not have long-lasting effects on the behaviour of $(\Delta_t)$. Overall, Assumptions \ref{assum:delta:filtration}, \ref{assum:moment:delta} and \ref{assum:moment:delta:diff} are quite common and are generally satisfied for the class of processes $(Z_t)$ (see \cite{xu2019multivariate} and \cite{guerrier2022robust} for a more detailed account on these assumptions). Under these assumptions, the consistency and asymptotic normality of $\sqrt{T}\left(\hat{\bm\nu}_{(Z)}-\bm\nu_{(Z)}\right)$ is ensured by \cite[Theorem 1]{xu2019multivariate}.

Unfortunately, although we have proven consistency of the WV estimator for the process $(S_t)$, we cannot make use of these conditions to prove the asymptotic normality of $\hat{\bm{\nu}}$ which denotes the estimator of WV when applied to the class of models of interest $(X_t)$. Indeed, Assumption \ref{assum:moment:delta:diff} cannot be satisfied for sinusoidal processes since, as discussed above, this assumption requires the process to have a ``short-range'' dependence property which is not reasonable for sinusoidal processes due to their intrinsic periodicity which, among others, does not allow the autocovariance structure to decrease at larger time lags. Nevertheless, the convergence rate of the WV estimator for sinusoidal processes ensures that the asymptotic normality of $\hat{\bm{\nu}}$ is still guaranteed as stated in Theorem \ref{thm:asymp_wv}. For the latter, we also define $\bm{\omega}_t := \left[\omega_{j,t}^{(Z)}\right]_{j=1, \hdots, J}$ as the vector of wavelet coefficients at time $t$ applied to the process $(Z_t)$ as well as the projection operator  $\mathcal{P}_{t}(\cdot) := \mathbb{E}\left[\cdot \mid \mathcal{F}_{t}\right]-\mathbb{E}\left[\cdot \mid \mathcal{F}_{t-1}\right].$ 
\begin{Theorem}\label{thm:asymp_wv}
For the Haar wavelet filter and under Assumptions \ref{assum:delta:filtration}, \ref{assum:moment:delta} and \ref{assum:moment:delta:diff}, we have
\begin{equation*}\label{wv_asymp}
   \sqrt{T}\left\{\hat{\bm{\nu}}-\bm{\nu}\left(\bm{\theta}_{0}\right)\right\} \stackrel{\mathcal{D}}{\rightarrow} \mathcal{N}(\mathbf{0}, \bm{V}),
\end{equation*}
where $\bm{V}=\mathbb{E}\left[\bm{D}_0 \bm{D}_0^{\top}\right]$ and $\bm{D}_0 \vcentcolon= \sum_{t=0}^{\infty} \mathcal{P}_{0}\left(\bm{\omega}_t\right)$.
\end{Theorem}
The proof of this theorem is omitted since it can be obtained directly by using Slutsky's Theorem when combining the result from Lemma \ref{lemma:sin:asymp} with that from Theorem 2.1 in \cite{guerrier2022robust} which proves asymptotic normality of $\hat{\bm{\nu}}$ under the above conditions. Indeed it can be noticed that, while the results hold for the WV estimator $\hat{\bm{\nu}}$, which is applied to the process $(X_t)$, the asymptotic covariance matrix is defined solely based on $\bm{\omega}_t$ which represent the wavelet coefficients from the decomposition of the process $(Z_t)$ that does not contain sinusoidal noise. As a consequence, for example, an estimator of the asymptotic covariance matrix $\bm{V}$ can be computed without taking into account the sinusoidal noise process, allowing to take advantage of existing results for this purpose. 
%
%

With Theorem \ref{thm:asymp_wv} we can now obtain the asymptotic distribution of the GMWM estimator $\hat{\bm{\theta}}$. To start, we need to consider the following additional assumptions.

\begin{Assumption}\label{assum:compact}
 $\bm\Theta$ is compact.
\end{Assumption}

\begin{Assumption}\label{assum:Omega}
If $\widehat{\bm\Omega}\in\real^{J\times J}$ is an  estimator of a definite-positive matrix $\bm\Omega$, then
 \begin{align*}
     \big\|\widehat{\bm\Omega} - \bm\Omega\big\|_S = o_p(1),
 \end{align*}
 where $\| \cdot \|_S$ denotes the spectral norm.
\end{Assumption}
\begin{Assumption}\label{assum:identif}
The function $\bm\nu(\bm\theta) = [\nu_j(\bm\theta)]_{j=1,\dots, J}$ identifies $\bt$,
in that for any $\bt_1,\bt_2\in\bT$  we have that $\nu_{j}(\bt_1)=\nu_{j}(\bt_2),$ for $j=1,\dots, J,$ implies $\bt_1=\bt_2$.
%
\end{Assumption}

Assumption \ref{assum:compact} is a common regularity condition which can eventually be replaced by a condition on the convexity of the parameter space that however can only be verified on a model-specific basis.  On the other hand, Assumption \ref{assum:Omega} is only required in case an estimator is chosen instead of any deterministic positive-definite matrix $\bm{\Omega}$. Indeed, if an estimator $\hat{\bm{\Omega}}$ is chosen, then this assumption requires this estimator to be consistent for the chosen positive-definite matrix $\bm{\Omega}$. Finally, an assumption that is also challenging to verify and is often assumed in practice is Assumption \ref{assum:identif}, which is equivalent to requiring that $\bm\nu(\bm\theta)$ be an injective function. In \cite{xu2019multivariate}, \cite{guerrier2020wavelet} and \cite{guerrier2022robust}, the validity of this assumption was discussed for different classes of composite processes which include combinations of, among others, white noise, random walk, quantization noise, drift and AR1 components. Hence, before stating the asymptotic properties of the GMWM, we add to these previous results by discussing the validity of this assumption when including a sinusoidal process in the class of composite processes defined in \eqref{eq:define:Xt}. As a first step, we verify this assumption when considering only one sinusoidal process through the following lemma.

\begin{lemma}
\label{lemma:ident:sinus}
For $J\geq 2$, the function $\bm{\nu}(\alpha, \beta)\vcentcolon=\left[\nu_j(\alpha, \beta) \right]_{j=1,\dots, J}$ identifies $(\alpha, \beta)$,
in that for any $(\alpha_1, \beta_1),(\alpha_2, \beta_2)\in\real_+\times(0,\pi]$,  we have that $\nu_{j}(\alpha_1, \beta_1)=\nu_{j}(\alpha_2, \beta_2),$ for $j=1,\dots J,$ implies $\quad(\alpha_1, \beta_1)=(\alpha_2, \beta_2)$.
%

\end{lemma}

The proof of Lemma \ref{lemma:ident:sinus} is given in Appendix \ref{appendix:proof:lemma:ident:sinus}. This result is important to confirm that the WV is informative with respect to this process, meaning that the information contained in the WV is sufficient to identify the parameters of the process in \eqref{sin_process_def}. We now try to extend this evidence towards the composite model \eqref{eq:define:Xt} for which we consider the lengths of the wavelet filters for each level $j$ (i.e., $\tau_j$) to live on a subset of the rational numbers representing the range of values containing those scales that would naturally arise in practice. More specifically, we assume that the WV scales are defined by $\tau \in \Lambda$ where $\Lambda \vcentcolon= \left[2, 2^J\right]\cap \mathbb{Q}$. This definition of the scales does not completely correspond to the ideal scenario for which we would like to provide supporting evidence for Assumption \ref{assum:identif}, but is an extension that allows to provide additional support to this assumption in the case of the process in \eqref{eq:define:Xt}. In this context we consider a model representative of the class defined in \eqref{eq:define:Xt} given by:
\begin{equation}
\label{eq.comp_model_ident}
     X_t \vcentcolon= W_t + Q_t +  Y_t + D_t + R_t +  S_t.
\end{equation}
%

%
%
%
%
%
In this case, the parameter space $\bm\Theta\subset\real^8$ is isomorphic to $\real_+^6\times\{(-1,0)\cup(0,1)\}\times(0,\pi]$ and, underlining that the notation $\bm{\nu}_{\tau}(\bm{\theta})$ refers to the WV over scales $\tau \in \Lambda$, the following lemma states the identifiability of the parameters of the model in \eqref{eq.comp_model_ident} within this continuous scale setting.

\begin{lemma}
\label{lemma:ident:sinus&others}

%

The function $\nu_\tau(\bm\theta)$ associated with the model in \eqref{eq.comp_model_ident} identifies $\bt$,
in that for any $\bt_1,\bt_2\in\bT\subset\real^8$,  we have that $ \nu_{\tau}(\bt_1)=\nu_{\tau}(\bt_2)$, for all $\tau\in\Lambda$ implies $\bt_1=\bt_2$.
%
\end{lemma}
The proof of this lemma is given in Appendix \ref{appendix:proof:lemma:ident:sinus&others}. The results of Lemmas \ref{lemma:ident:sinus} and \ref{lemma:ident:sinus&others} are therefore helpful in supporting the validity of Assumption \ref{assum:identif} when adding sinusoidal processes to the modelling framework. More specifically, if only considering the sinusoidal process $(S_t)$ in the model in \eqref{eq.comp_model_ident}, it is obvious that the identifiability through $\bm\nu(\bm\theta)$, given in Lemma \ref{lemma:ident:sinus}, implies Lemma \ref{lemma:ident:sinus&others} but the converse is not necessarily true. Therefore Lemma \ref{lemma:ident:sinus&others} is useful but does not imply the identifiability of the general class of models defined in \eqref{eq:define:Xt}.

Following the results in \cite{xu2019multivariate} and \cite{guerrier2022robust}, the GMWM estimator $\hat{\bm{\theta}}$ is consistent for the class of models in \eqref{eq:define:Xt} under Assumptions \ref{assum:delta:filtration} to \ref{assum:identif}. In order to obtain its asymptotic normality, we need to consider two additional assumptions defined below.



%
\begin{Assumption}\label{assum:convex}
$\bm\Theta\subset \real^p$ is convex and $\bm\theta_0\in\bm\Theta$ is an interior point.
\end{Assumption}

\begin{Assumption}\label{assum:dervi:nu}
The derivative $\bm{A}(\bm\theta_0)\vcentcolon=\dfrac{\partial}{\partial \bm\theta^T}\bm\nu(\bm\theta)\Big|_{\bm\theta = \bm\theta_0}$ is such that 
\begin{align*}
  \bm B(\bm\theta_0) \vcentcolon= \bm{A}(\bm\theta_0)^T\bm\Omega\bm{A}(\bm\theta_0),
\end{align*}
is non-singular.
\end{Assumption}

Assumption \ref{assum:convex} is a regularity condition that allows us to make use of the mean value theorem which can nevertheless be quite restrictive, for example in the case where the assumed model overfits the data. Indeed in the latter case some components of the parameter vector $\bm\theta_0$ may lie on the border of $\bm\Theta$ if, for example, some variance parameters are equal to zero. Assumption \ref{assum:dervi:nu} on the other hand simply enables us to define the asymptotic covariance matrix of $\hat{\bm\theta}$. Denoting $\bm M\in\real^{J\times p}$ and $\bm N\in\real^{J\times J}$, we define the operator $\bm M\boxtimes \bm N\vcentcolon=\bm M\bm N\bm M^T$ which allows us to state the following theorem and delivers the final result on $\hat{\bm{\theta}}$.
\begin{Theorem}\label{thm:asymp:norm}
Under Assumptions \ref{assum:delta:filtration} to \ref{assum:dervi:nu},  we have that 
\begin{align*}
    \sqrt{T}\left(\hat{\bm\theta}-\bm\theta_0\right)\stackrel{\mathcal{D}}{\longrightarrow}\mathcal{N}(\bm{0},\bm\Xi),
\end{align*}
where $\bm\Xi \vcentcolon= \{\bm B(\bm\theta_0)^{-1} \bm{A}(\bm\theta_0)^{T}\bm\Omega\} \boxtimes \bm V$ and $\bm V$ is given in \eqref{wv_asymp}.
\end{Theorem}
As for consistency, this theorem follows directly from the results in \cite{xu2019multivariate} and \cite{guerrier2022robust}. In brief, based on Assumption \ref{assum:convex} we can use the mean value theorem on the GMWM objective function in \eqref{def:GMWM} around the true value $\bm{\theta}_0$ and, based on the consistency of $\hat{\bm{\theta}}$, it is possible to show convergence of the different quantities defined by this expansion (including that of the derivative $\bm{A}(\bm{\theta})$) towards their theoretical values which define the asymptotic covariance matrix. 

\begin{Remark}
It can be noticed how all quantities that define the asymptotic covariance $\bm\Xi$ depend on the parameter $\bm{\theta}_0$ with the exception of $\bm{\Omega}$ and $\bm{V}$. While the quantities that depend on $\bm{\theta}_0$ can be estimated by plugging in the consistent estimator $\hat{\bm{\theta}}$, the matrix $\bm{\Omega}$ is chosen by the user while $\bm{V}$ has to be estimated. Given the results in this work, the asymptotic covariance matrix $\bm{V}$ can be estimated using the proposals in \cite{xu2019multivariate} and \cite{guerrier2022robust} without considering the presence of vibration noise. Moreover, making the choice $\bm\Omega \vcentcolon = \bm V^{-1}$ delivers the most asymptotically efficient GMWM estimator $\hat{\bm\theta}$ (see \citealp{hansen1982large}). Indeed, for this particular choice of the weight matrix the expression of the GMWM covariance matrix simplifies to $\bm\Xi = \{\bm{A}(\bm\theta_0)^{T}\boxtimes \bm V^{-1}\}^{-1} = \bm B(\bm\theta_0)^{-1}$. This expression is actually the same as the one obtained in the just-identified case, i.e., when $J=p$. In the latter case, we also have  $\bm\Xi = B(\bm\theta_0)^{-1}$ independently of the choice of $\bm\Omega$.
\end{Remark}

%

\section{Simulation Studies}
\label{section_simu}

In this section we present different simulation studies to investigate the performance of the GMWM when considering various models included in the general class defined in \eqref{eq:define:Xt}, specifically those characterized by the presence of sinusoidal processes which we consider as a nuisance noise. In particular, although these cannot be considered a proof for parameter identifiability, the simulations also aim at understanding to what extent the GMWM is able to identify the parameters of models which include sums of sinusoidal noises and other latent processes. In addition, we want to understand the loss of statistical efficiency of the GMWM with respect to more computationally-demanding likelihood-based approaches (and also how much computational gain is achieved when using the GMWM). In all cases we simulate $500$ signals from each model and choose parameter values that are consistent with those that are commonly identified for the stochastic errors of measurement devices such as IMUs (see e.g., \cite{titterton2004strapdown, el2007analysis, guerrier2016theoretical} and the applied case study in Section  \ref{section_application}). The parameter values for each of the simulations presented in this section are reported in Appendix~\ref{appendix:simu_param_values}.

\begin{figure}
    \centering
    
    \includegraphics[scale=.85]{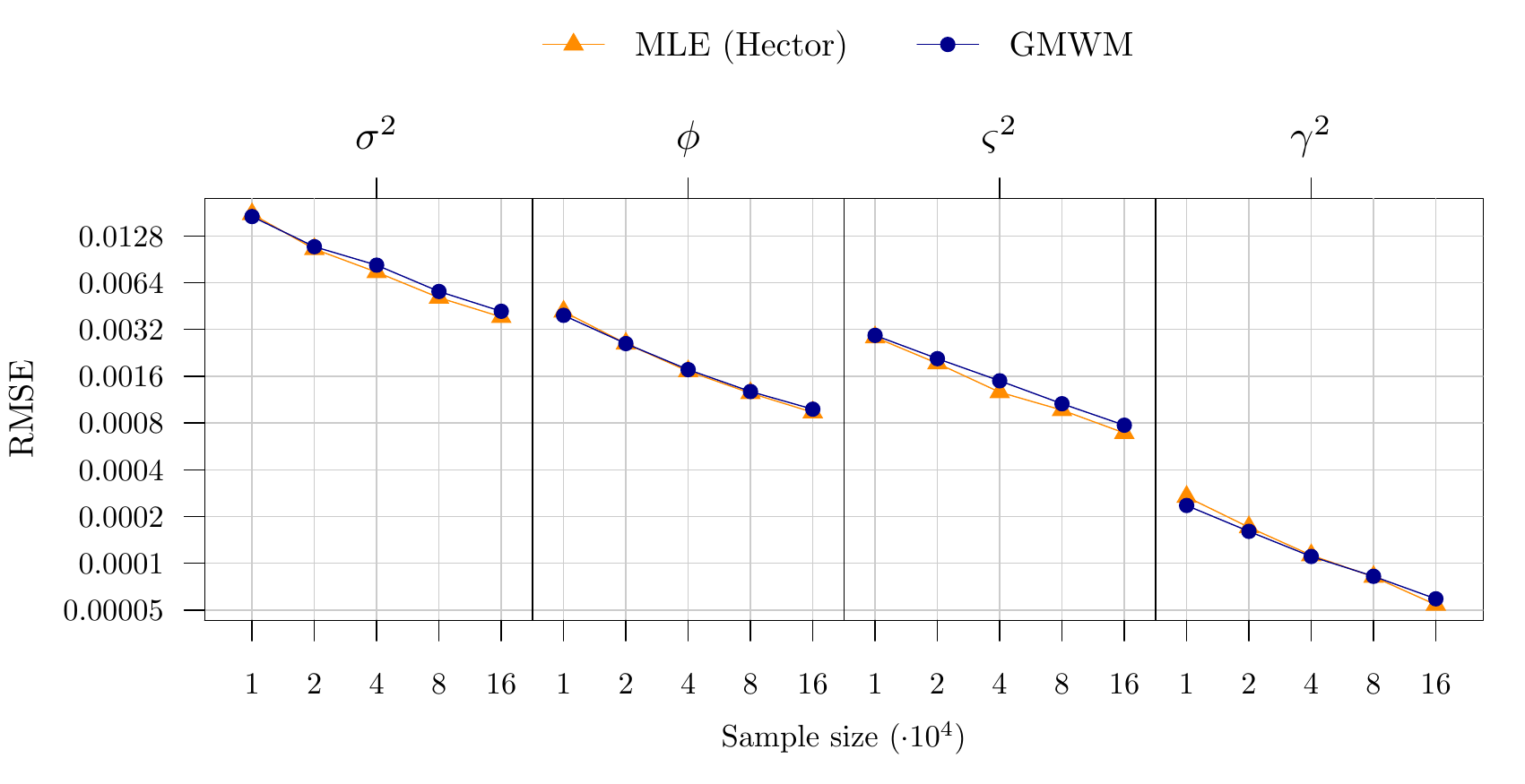}
    \caption{RMSE of the estimated parameters of Simulation~1 for the MLE (orange line) and the GMWM (blue line) for sample sizes  $T=T_i\cdot10^4$ with $\{T_1,\dots,T_5\}=\{1, 2, 4, 8, 16\}$}.
    \label{fig:rmse_simu_1}
\end{figure}

\begin{figure}
    \centering
    \includegraphics[scale=.9]{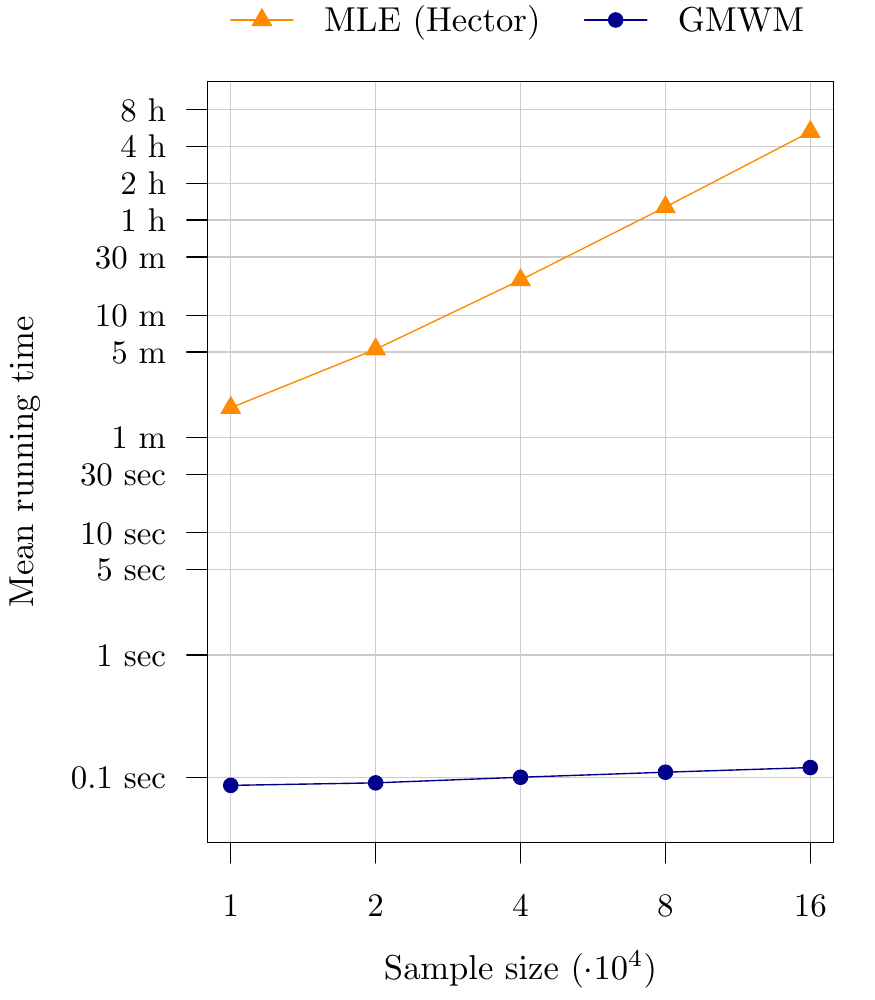}
    \caption{Mean running time of the MLE (orange line) and the GMWM (blue line) in Simulation~1 for sample sizes $T=T_i\cdot10^4$ with $\{T_1,\dots,T_5\}=\{1, 2, 4, 8, 16\}$}.
    \label{fig:running_time_mle_gmwm}
\end{figure}

For the first study (which we refer to as Simulation~1), we consider a model represented by the sum of a WN, a RW, an AR1 and a single sinusoidal process therefore requiring the estimation of six parameters in total (see Appendix \ref{appendix:simu_param_values} for values) considering the parametrization in \eqref{sin_process_def}. We use this model to compare the statistical and computational performance of the GMWM with respect to the MLE implemented in the open-source software \texttt{Hector} (v2.0) which represents the fastest available implementation of the MLE for these latent models (see \citealp{Bos2008, bos2013fast}). It must be noted that, for the latter approach the vibration noise is purely deterministic and assumes that the frequency is known, leaving amplitude and phase to be estimated. Due to these different parametrizations, the two approaches cannot be compared in terms of estimation of the sinusoidal process parameters, but only in terms of estimation of the processes of true interest (i.e., WN, RW and AR1). For this purpose, considering commonly large signals recorded by high-frequency measurement devices, we generate signals of five different lengths, i.e., $T=T_i\cdot10^4$ with $\{T_1,\dots,T_5\}=\{1, 2, 4, 8, 16\}$, and compute the Root Mean Squared Error (RMSE) as well as the average running time for both approaches. This information is represented in Figures~\ref{fig:rmse_simu_1} and \ref{fig:running_time_mle_gmwm} respectively where, for both methods, we removed results that were affected by convergence problems for the MLE (specifically for smaller sample sizes) in order to make fair comparisons. We denote the parameters as: $\sigma^2 \in \real_+$ (WN variance); $\phi \in (-1,0)\cup(0,1)$ and $\zeta^2 \in \real_+$ (AR1 autoregressive and innovation variance parameters respectively) and $\gamma^2 \in \real_+$ (RW innovation variance). As can be observed in Figure~\ref{fig:rmse_simu_1}, for all parameters of interest both methods have an RMSE that decreases with the sample size thereby supporting consistency of the GMWM and the MLE in this setting. Moreover it can be seen how, with few marginal differences, the RMSE of both methods appear to be extremely close to each other suggesting that the potential loss of statistical efficiency of the GMWM with respect to the MLE is almost negligible in sample sizes of relevance for the considered applications. This conclusion needs to be evaluated jointly with the results presented in Figure~\ref{fig:running_time_mle_gmwm}: as it can be observed, the average MLE running time ranges from less than $2$ seconds (for $T = 10^4$) up to more than $5$ hours (for $T = 16 \cdot 10^4$) while the GMWM consistently runs in less than half a second for all sample sizes considered. This implies that, with comparable performance in terms of RMSE, the GMWM is at least $12 \cdot 10^2$ times faster on average than the MLE in the considered sample size settings ($15 \cdot 10^4$ faster for the largest sample size). It must also be noted that, for smaller sample sizes the MLE suffers from convergence issues which is not the case for the GMWM.

For the next simulation settings, we consider more complex models and larger sample sizes which are often observed in real measurement error signals. In these settings the MLE can become more numerically unstable and remains computationally demanding for these sample sizes. Therefore, considering the comparison made in Simulation~1, in the next simulations we only verify the performance of the GMWM and, as a result, also focus on the estimation of the parameters of the sinusoidal process put forward in \eqref{sin_process_def}. More in detail, we first consider a  model defined as the sum of two AR1 processes, with parameters $|\phi_i| \in (0,1)$ and $\zeta_i^2 \in \real_+$, and two sinusoidal processes, with parameters $\alpha_i \in \real_+$ and $\beta_i \in (0,\pi]$, for $i = 1, 2$ (we refer to this setting as Simulation~2) and then a model composed of an AR1, a RW, a WN and a sinusoidal process, all with parameter notations consistent with the previous simulations (we refer to this setting as Simulation~3). For these two simulations we consider a sample size of $1 \cdot 10^7$ and, as mentioned previously, simulate $500$ time series for both simulations with parameter values in the range of those found in practical applications such as those discussed in \cite{guerrier2013wavelet} and \cite{stebler2014generalized} (see Appendix \ref{appendix:simu_param_values} for values). In particular, due to the high-frequency of the measurements, the autoregressive parameters $\phi$ for example can be close to unit value (i.e., close to a RW process) creating additional numerical convergence issues of the MLE. 

\begin{figure}
    \centering
    \includegraphics[scale=.665]{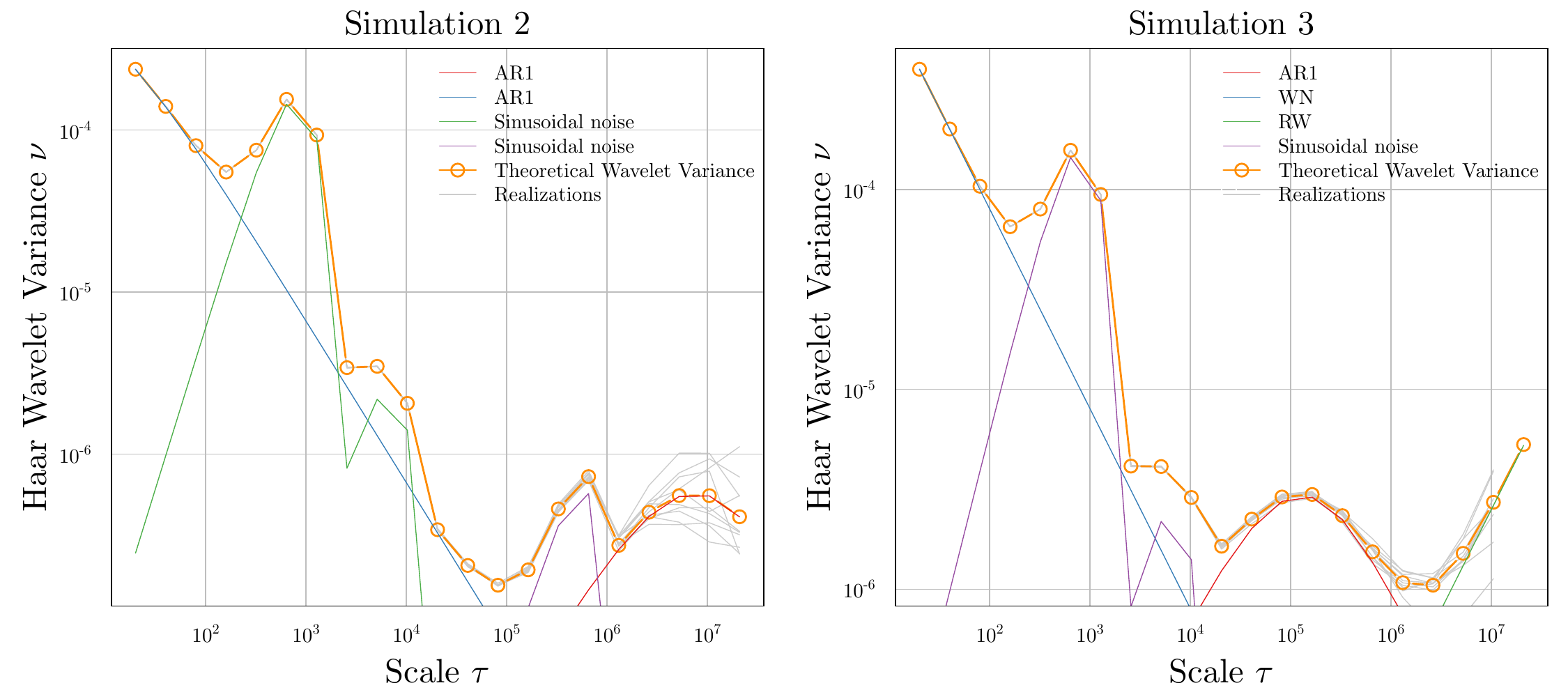}
    \caption{Theoretical wavelet variance $\bm{\nu}(\bm{\theta})$ (orange line) of the settings considered in Simulation~2 and Simulation~3. The light grey lines represent WV estimates $\hat{\bm{\nu}}$ from 10 different realizations of the respective models. The other lines represent the theoretical WV of the individual processes contributing to the models.}
    \label{fig:wavelet_variance_simu_2_3}
\end{figure}

\begin{figure}
    \centering
    \includegraphics[scale=.68]{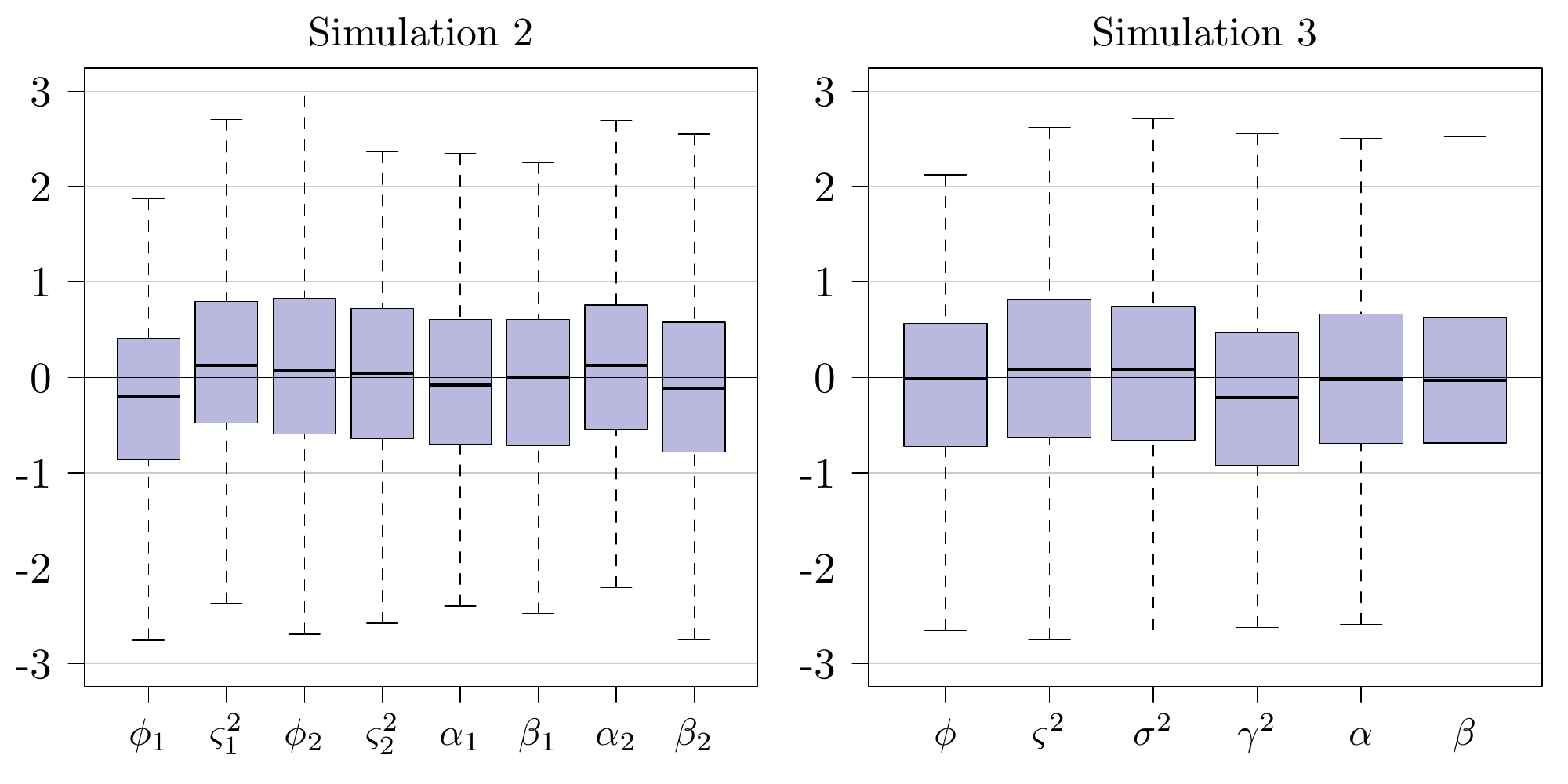}
    \caption{Empirical distribution of GMWM parameter estimates for Simulations 2 and 3. The true parameter values were subtracted from each boxplot and all distributions were standardized by their respective empirical variances.}
    \label{fig:boxplot_simu_gmwm_2}
\end{figure}

A representation of these two models is given through a WV plot in Figure~\ref{fig:wavelet_variance_simu_2_3} where it can be seen that the theoretical WV $\bm{\nu}(\bm{\theta})$ of each model (orange line) is composed from the contribution of the WV of the individual processes, generating realizations of empirical WV estimates $\hat{\bm{\nu}}$ (light grey lines) that closely follow it. Hence, as explained in the previous section, the GMWM observes the grey line $\hat{\bm{\nu}}$ and aims at finding the parameter vector $\bm{\theta} \in \bT$ that allows the implied theoretical WV $\bm{\nu}(\bm{\theta})$ (orange line) to be as close as possible to this empirical WV in $L_2$-norm (weighted by $\bm{\Omega}$). The empirical distributions of the estimated parameter values are represented through the boxplots in Figure \ref{fig:boxplot_simu_gmwm_2}. We subtract the true parameter values from these distributions (hence all boxplots should be roughly centered around zero if the GMWM is correctly estimating these parameters) and standardize them via their respective empirical standard deviations to compare them all on the same scale. We consider this re-scaling since we cannot compare these distributions to those of other estimators, therefore we are mainly interested in consistency rather than efficiency of the GMWM, which however can be observed in the boxplots in Appendix \ref{appendix:simu_param_values} as well as partially studied through the RMSE in smaller sample sizes in Simulation~1 (see Figure \ref{fig:rmse_simu_1}). As highlighted by the two plots in Figure \ref{fig:boxplot_simu_gmwm_2}, the GMWM appears to correctly target the true values of the parameters of the models considered in both simulations, including those of the sinusoidal processes put forward in this work thereby supporting the theoretical results in Section \ref{sec.method}.


\section{Case Study: Inertial Sensor Calibration}
\label{section_application}

To study the advantage of the proposed approach for real-world applications, we consider stochastic measurement error data (of size $T = 2,879,999$) collected from controlled IMU calibration sessions. More specifically, to study the impact of vibration noise, the stochastic measurement errors of the same z-axis gyroscope of a low cost MEMS IMU were measured at 200 Hz in two controlled settings: (i) a static setting and (ii) a rotating setting where the IMU is placed on rotating table at a fixed rotation-speed of $200$ deg/sec. The latter setting is used to deliver the possible vibrations that can often corrupt measurement devices during their calibration phases. The intuition for this application is that the stochastic model for the measurement error in static settings constitutes the basic error specific to the measurement device which we are interested in, however this model becomes more complex under rotation through the addition of one or more vibration noise components which constitute disturbances to the true stochastic error. Hence, we would like to understand how the basic measurement error model changes when the additional vibration noise is taken into account and, as a consequence, what are the impacts of this change when employing the estimated models (with and without considering vibration noise) in navigation settings.

\begin{figure}
    \centering
    \includegraphics[scale=.625]{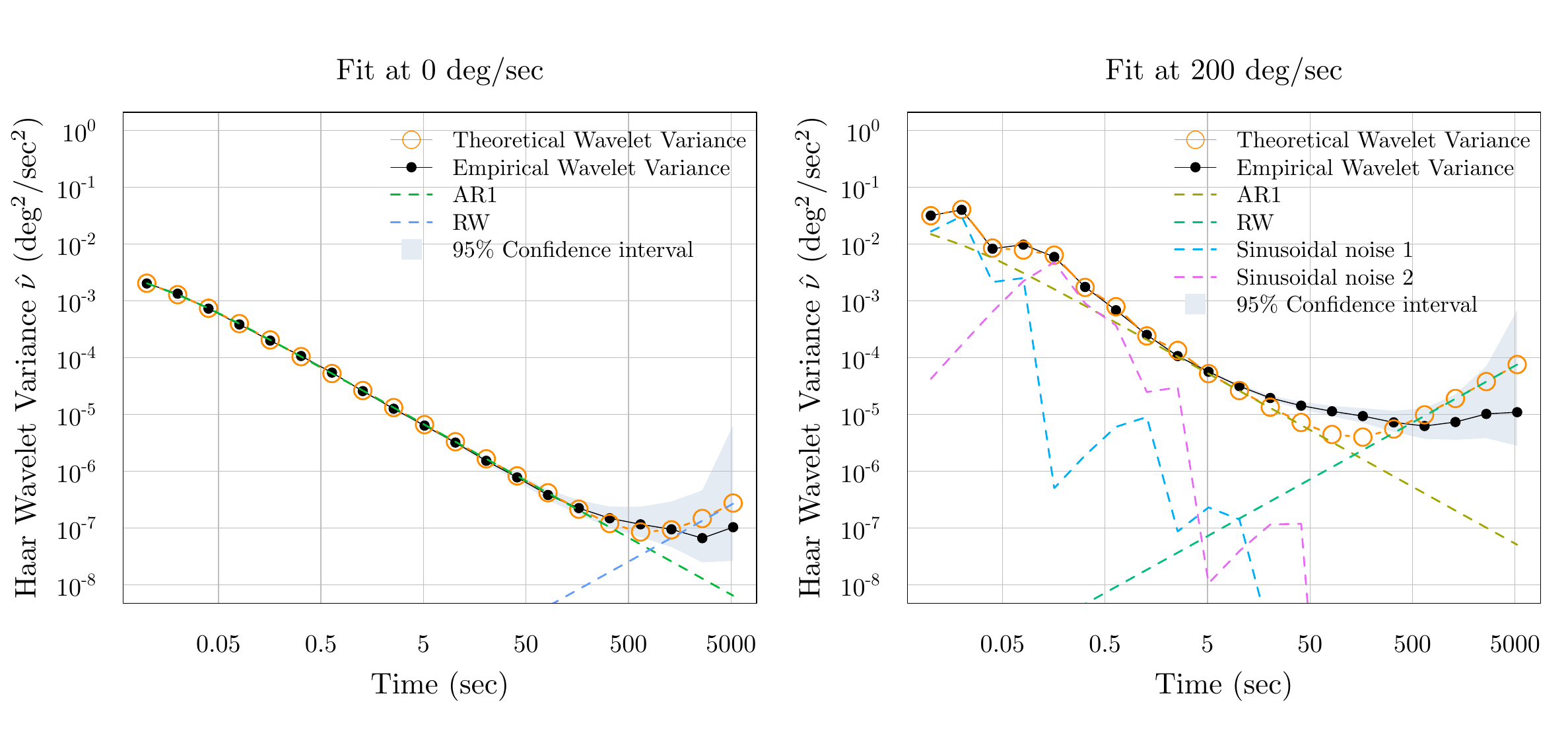}
    \caption{Empirical WV (black line) of the MEMS IMU in static conditions (left plot) and under $200$ deg/sec rotation (right plot). The WV implied by the respective models estimated via the GMWM is represented by the orange line while the contribution of each underlying process to these models is represented through lines of other colors.}
    \label{fig:application_both_estimated_fit}
\end{figure}

The basic model for the IMU was chosen via a visual representation of the WV (black line) shown in the left plot of Figure \ref{fig:application_both_estimated_fit}. Comparing different models, the composition of an AR1 and a RW process (orange line) appeared to best fit the observed empirical WV on the measurement error in static settings. Indeed, the theoretical WV implied by the GMWM estimates of this model (AR1 and RW) appears to closely follow the empirical WV. Having identified the basic model for the stochastic error of this IMU, we considered the empirical WV of the same IMU under the above-mentioned rotation setting which can be observed in the right plot of Figure \ref{fig:application_both_estimated_fit}. In the latter setting the assumption is that the structure of the basic model remains the same but is complemented with additional vibration noise in the form of one or more sinusoidal processes which themselves can have an impact on the parameter values of the basic model. Hence, preserving this basic model structure, the addition of two sinusoidal processes appear to better fit the WV observed under rotation as hinted by the right plot of Figure \ref{fig:application_both_estimated_fit}. Indeed, the postulated parametrization in \eqref{sin_process_def} appears to well describe the additional structure in the empirical WV observed under rotation, therefore supporting the usefulness of the approach put forward in this work. More in detail, Table \ref{table:imu_parameters} reports the parameters of the basic model (i.e., AR1 and RW parameters) when estimated in static and rotating conditions (including sinusoidal processes) respectively. It can be seen how all the estimated parameters of the basic model differ significantly when considered in the two settings, highlighting how the presence of vibrations due to rotation affects the values of this model's parameters.

\begin{table}[ht]
\resizebox{\columnwidth}{!}{%
\begin{tabular}{c|cccc}
\multirow{2}{*}{${}_{\mathrm{Parameter}}$ $\mkern-6mu\setminus\mkern-6mu {}^{\hspace{1mm}  \mathrm{Rotation} \hspace{.7mm} \mathrm{rate}}$} &  \multicolumn{2}{c}{0 \hspace{0.3em} deg/sec} & \multicolumn{2}{c}{200 \hspace{0.3em} deg/sec}\\ 
& Estimate & CI(95 \%) &Estimate & CI(95 \%) \\
 \hline 
\rule{0pt}{2.5ex}    
$\phi$ & $1.63 \cdot 10^{-1\phantom{0}}$ & $(1.62 \cdot 10^{-1\phantom{0}}$; $1.64 \cdot 10^{-1\phantom{0}}$) & $1.85 \cdot 10^{-1\phantom{0}}$ & $(1.81 \cdot 10^{-1\phantom{0}} ; 1.89 \cdot 10^{-1})$ \\
$\varsigma^2$ & $4.78 \cdot 10^{-3\phantom{0}}$ & $(4.77 \cdot 10^{-3\phantom{0}};  4.78 \cdot 10^{-3\phantom{0}} )$ & $3.56 \cdot 10^{-2\phantom{0}}$ & $(3.55 \cdot 10^{-2\phantom{0}} ; 3.57 \cdot 10^{-2}) $ \\
$\gamma^2$ & $3.09 \cdot 10^{-12}$ & $(9.01 \cdot 10^{-13}; 5.52 \cdot 10^{-12})$ &$ 8.69 \cdot 10^{-10}$ & $(5.88 \cdot 10^{-10} ; 1.12 \cdot 10^{-9}) $

\end{tabular}
}
    \caption{Estimated parameters and $95\%$ parametric bootstrap confidence intervals for the error signal collected at $0$ deg/sec (i.e., corresponding to the first and second column respectively) and for the error signal collected at $200$ deg/sec (i.e., corresponding to the third and fourth column respectively).}
    \label{table:imu_parameters}
\end{table}

\begin{figure}
    \centering
    \includegraphics[scale=.9]{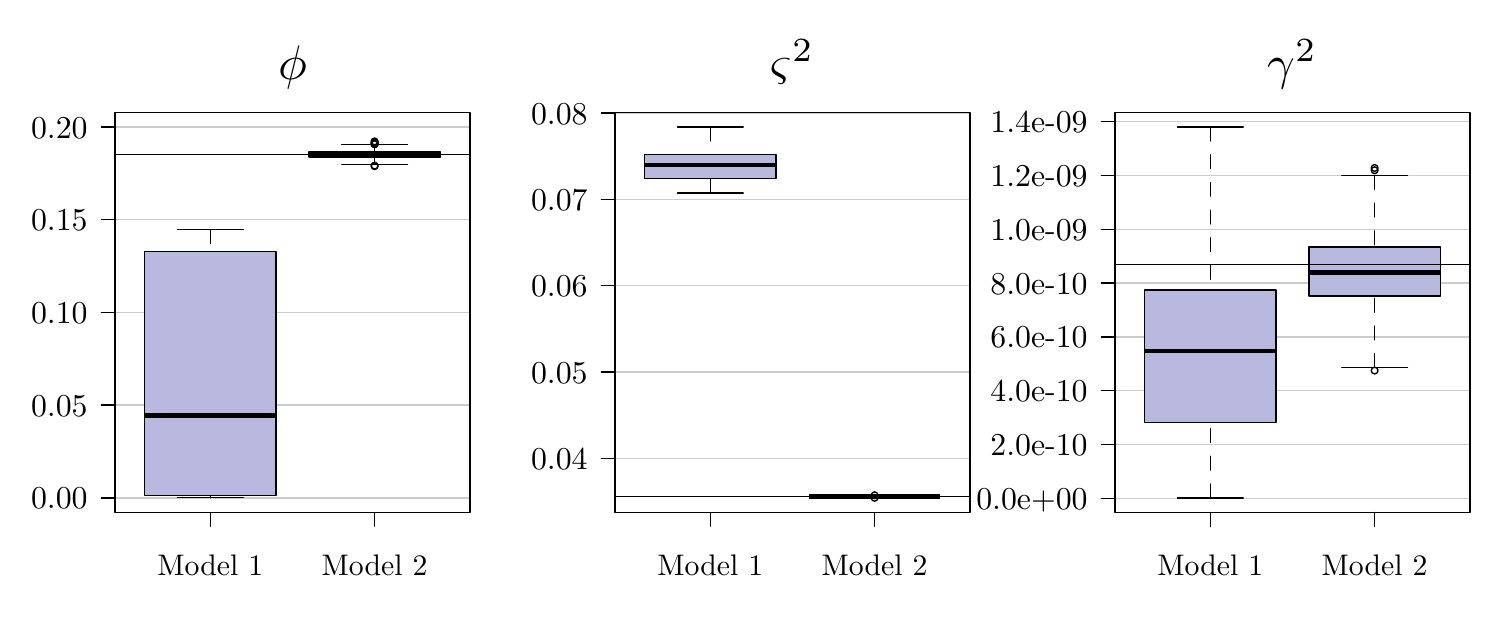}
    \caption{Empirical distribution of the GMWM parameter estimates under Model 1 (misspecified) and Model 2 (correctly specified) represented in the left and right boxplot in each plot respectively.}
    \label{fig:boxplot_estimated_param_navigation_simu}
\end{figure}

To date, as described earlier in this work, the inclusion of a stochastic process to account for noise induced by the vibration of measurement devices has not been fully addressed, if not through deterministic models for which certain parameters need to be known in advance (and with important numerical/computational issues for existing methods). Hence, the common approach to this kind of setting is to approximate these error signals through the standard models (e.g., AR1 and RW) under static scenarios which however can often be affected by different sources of vibration. Considering this, we will study the extent of the bias induced by excluding the sinusoidal processes, that are present during the calibration phase, through a simulation study based on the parameter estimates from the MEMS IMU under rotation in Table \ref{table:imu_parameters} (the estimated parameters of the sinusoidal processes are given in Appendix \ref{appendix:simu_param_values}). Therefore we will simulate from the model that includes the two sinusoidal processes and then, each time, use the GMWM to estimate a misspecified model with only an AR1 and a RW process (Model 1) as well as a correctly specified model which also includes the two sinusoidal processes (Model 2). We focus on how much model-misspecification in this scenario impacts the estimates of the basic model of interest composed of the AR1 and RW processes. The results of this simulation, based on the real estimates from the considered MEMS IMU data, are represented in the boxplots of Figure \ref{fig:boxplot_estimated_param_navigation_simu}. As expected, from these boxplots it is clear to what extent the model misspecification can have significantly negative impacts both in terms of bias as well as in terms of variance, especially with respect to the AR1 parameters.

The previous simulation is a simple proof-of-concept of the intuitive fact that not accounting for the vibration noise, when this is actually present, can severely impact estimation for the other (basic) model parameters. However, this does not necessarily give the idea of the impact that this problem may have in real-world applications. For this reason, we translate the above simulation setting to a navigation scenario where we take the estimated model under rotation (i.e., Model 2 which includes the two sinusoidal processes) and generate stochastic error signals that we will add to an arbitrarily determined and fixed navigation path defined by a trajectory and an altitude profile (this is known as an \textit{emulation} study). In the latter scenario we can imagine that this path describes the movement of an aerial vehicle (e.g., a drone) and, for this emulation, we imagine that this vehicle is guided by an integrated navigation system composed of a Global Positioning System (GPS) and IMUs such as the one considered in this section. In these systems the most accurate measurements are given by the GPS while the IMUs are used mainly to update navigation estimates between GPS measurements and also for uncertainty quantification. Unfortunately, GPS signals can be corrupted or be absent in different situations and, as a consequence, the IMUs are employed in so-called ``coasting mode'' to provide the navigation solutions without the GPS. In this case, and in general, it is extremely important for the navigation filter associated to these IMUs, usually an Extended Kalman Filter (EKF), to be programmed with precise estimates of the stochastic error signals that characterize them so that they can be removed from their measurements and obtain more accurate navigation solutions for the vehicle. We rely on the framework presented in \cite{Cucci2022} to perform this emulation study and assess the impact on navigation performances when considering or not the sinusoidal perturbations in the stochastic calibration procedure.

\begin{figure}
    \centering
    \includegraphics[scale=.8]{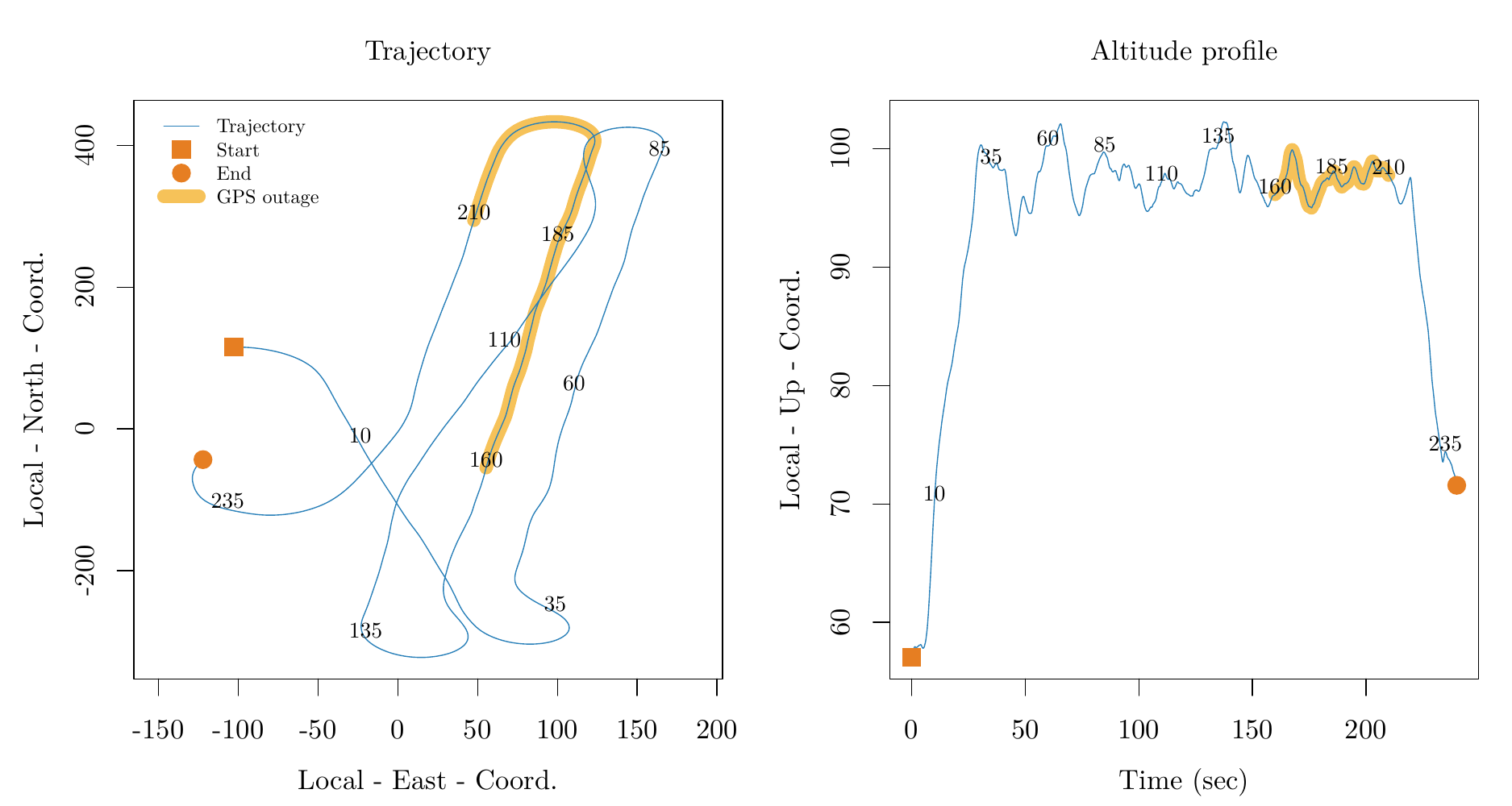}
    \caption{Trajectory and altitude profile (blue lines) of the vehicle considered in the emulation study over $235$ sec. The highlighted part of these paths (in yellow) represent the portion in which we mimic the GPS outage (from $160$ sec to $210$ sec).}
    \label{fig:trajectory}
\end{figure}

\begin{figure}
    \centering
    \includegraphics[scale=.75]{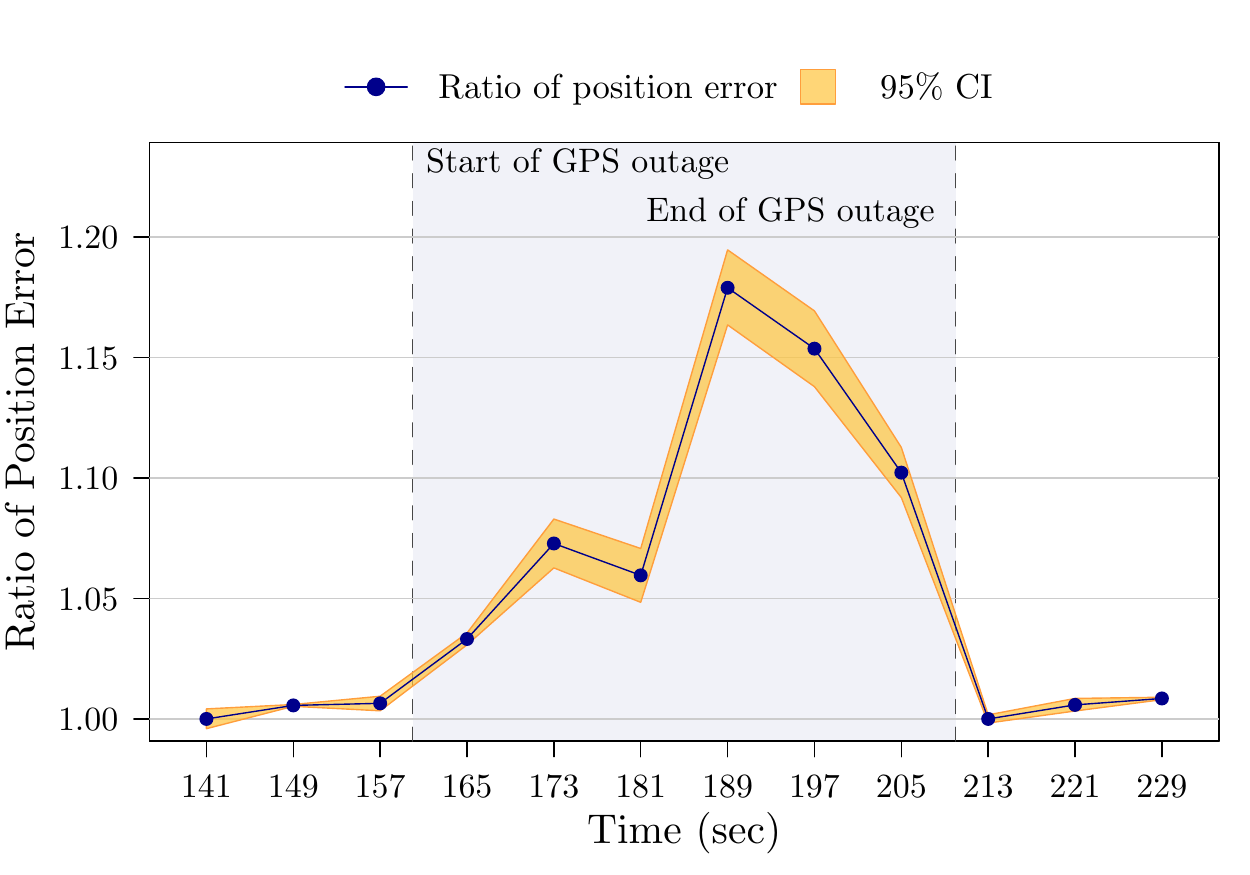}
    \caption{Mean ratio of position errors for EKF based on $\hat{\bm{\theta}}_1$ over those for EKF based on $\hat{\bm{\theta}}_2$. Yellow bands represent the 95\% confidence intervals for the mean ratio.}
    \label{fig:line_chart_position_error}
\end{figure}


Given this setting, let us perform an emulation study where we determine a ground-truth trajectory and altitude profile for a period of 235 seconds during which we mimic a GPS outage for 50 seconds (from 160 seconds to 210 seconds) as represented in Figure \ref{fig:trajectory}. More in detail, we assume that the GPS is measuring at 1 Hz while the IMU is measuring at 100 Hz. Hence, let us study how much the navigation solutions, more specifically \textit{position} solutions, are affected by the approach taken in the previous simulation which highlights the effects of not accounting for vibration noise (see Figure \ref{fig:boxplot_estimated_param_navigation_simu}). To do so, let us first define $\bm{\lambda} \in\real_+^4 \times\{(-1,0)\cup(0,1)\}\times(0,\pi]^2$ as the parameter vector containing the GMWM estimates for Model 2 taken on the IMU under rotation (therefore containing AR1, RW and two sinusoidal processes) while we will define $\bm{\theta}\in\real_+^2 \times\{(-1,0)\cup(0,1)\}$ as the general parameter notation of the \textit{basic} model structure consisting solely in the sum of an AR1 and a RW process (this is indeed the structure of Model 1). Given this, we can use $\bm{\theta}^\star$ to represent the elements of $\bm{\lambda}$ that correspond to the parameters of this basic model, hence without the parameters of the sinusoidal processes. Using this notation, we take the following approach for the $b^{th}$ iteration (out of 500): (i) we simulate a stochastic error signal from Model 2 based on $\bm{\lambda}$ and of the same length as the original data (i.e., $T = 2,879,999$) that we refer to as $(x_t^{(b)})$; (ii) we use $(x_t^{(b)})$ to estimate Model 1 (misspecified) and Model 2 (correctly specified) using the GMWM and denote the associated basic model parameters as $\hat{\bm{\theta}}_1^{(b)}$ and $\hat{\bm{\theta}}_2^{(b)}$ respectively (hence we discard the estimates of the parameters of the sinusoidal processes in Model 2); (iii) we simulate an additional stochastic error signal but this time from the basic model (AR1 and RW) based on $\bm{\theta}^\star$, and add this error signal to the previously mentioned navigation paths represented in Figure \ref{fig:trajectory} (blue lines); (iv) we use an EKF based on $\hat{\bm{\theta}}_1^{(b)}$ and $\hat{\bm{\theta}}_2^{(b)}$ respectively to estimate the position of the vehicle when the GPS is available as well as when there is an outage. Following this, at each time point we have a position estimate using the basic model parameters estimated under Model 1 and Model 2 respectively. Hence it is possible to compute a position error ratio between the two solutions throughout the entire emulated path. The position error is defined as the average over the emulated trajectories of the $\ell_2$-norm of the position error over the three axes defined as $ \Delta r_t= \hat{r}_t-r_t$, where $r_t$ denotes the true position at time $t$ and $ \hat{r}_t$ is the estimated position at time $t$. The results of this ratio (position error based on $\hat{\bm{\theta}}_1$ over position error based on $\hat{\bm{\theta}}_2$) are given in Figure \ref{fig:line_chart_position_error} where we represent the mean position ratio along with its corresponding 95\% confidence intervals. It can be observed how this ratio is almost always above one when the GPS signal is available, indicating that the model that accounts for vibration has a slightly lower position error in-between GPS measurements. This observation is confirmed when the GPS outage occurs (grey area): the ratio drastically and steadily increases up to $1.1789$ indicating a rapidly deteriorating position error for the misspecified navigation filter, only to return towards one when the GPS signal is available again. While this emulation study substantially confirms the intuitive bias that can be induced by an omission of stochastic disturbances during calibration (i.e., vibration noise), it also provides more insight to the real-life impacts that such an omission can entail. In this example we can observe how much a misspecified navigation filter, based on a model that does not account for vibration noise during the calibration phase, can severely affect navigation precision of (autonomous) vehicles. Indeed, a difference in position error of $15$\% is considered large and in our study it actually goes beyond $17$\%. If one wanted to correct an error of such magnitude, this would generally require much more accurate and expensive equipment. This study therefore highlights how a significant improvement in navigation can be achieved when accounting for vibration noise  with important advantages for  many applications going from ground-vehicles to drones for aerial-mapping and rescue-searches. In these applications it is of essence to track their positions in situations where GPS signals are often absent, thereby highlighting how much this sizeable navigation error can affect the successful outcome of these tasks. 

\section{Conclusions}
\label{section_conclusion}

The proposal of a stochastic parametrization of wave functions to account for vibration noise can be helpful in many applications ranging from engineering to natural sciences. In particular, modeling this noise while considering the presence of several other stochastic processes in large signals is a computational and/or numerical challenge for standard approaches. The derivation of theoretical forms for the WV and the study of its properties in the context of sinusoidal processes has allowed to extend the flexibility of the GMWM modeling framework which can account for more complex features in signals in a computationally efficient and numerically stable manner, thereby greatly improving the precision of measurement devices which was the specific focus of this work. More broadly though, this methodology can be applied in a wide range of domains where periodic signals are observed and need to be taken in account when performing statistical modeling and inference on time series data. For example, our approach could be applied in the context of modeling daily position time series from Global Navigation Satellite Systems where periodic signals often need to be considered jointly with other deterministic and stochastic signals as highlighted in \cite{cucci2023generalized}.

\newpage

\bibliographystyle{chicago.bst}
\bibliography{bibli}

\newpage

\appendix
\centerline{\Large\sc Appendices}

\section{Proof of Lemma \ref{lemma:sin:asymp}}
\label{appendix:proof:lemma:sin:asymp}%

In this proof, $\hat{\nu}_j$ correspond to the wavelet variance estimator associated with a single sinusoidal process, i.e., we have, by definition,
\begin{align*}
    \hat{\nu}_j  & = \frac{1}{M_{j}}\sum_{t=\tau_{j}}^{T} \left(\sum_{l=0}^{\tau_{j}-1} h_{j,l}S_{t-l}\right)^2.
\end{align*}
and $M_{j}   = T - \tau_{j}$.
With Wavelet coefficient     $h_{j,l}=\left\{
        \begin{array}{lcl}
            \phantom{-} \frac{1}{\tau_{j}}  &\text{for}  &l\leqslant\frac{\tau_{j}}{2} - 1\\
            -\frac{1}{\tau_{j}} &\text{for}  &l\geqslant\frac{\tau_{j}}{2} 
        \end{array} 
    \right.,$ 
we then have \begin{align*}
    \hat{\nu}_j  & = \frac{1}{\tau_{j}^{2}}\frac{1}{M_{j}}\sum_{t=\tau_{j}}^{T}\left(\sum_{l=0}^{\frac{\tau_{j}}{2}-1} S_{t-l} - S_{t-\frac{\tau_{j}}             {2}-l}\right)^2.
\end{align*}
%
            
%
Plugging $S_{t} = \alpha \sin\left(\beta t + u\right)$, where $\alpha\!\in\!\real_+$, $\beta\in(0,\pi]$ and $u \in (0, 2\pi)$, into the above equality, we define 
\begin{align*}
    {\nu}_{j,u}  & \vcentcolon= \frac{\alpha^2}{\tau_{j}^{2}}\frac{1}{M_{j}}\sum_{t=\tau_{j}}^{T}\left[\sum_{l=0}^{\frac{\tau_{j}}{2}-1}  \sin\left\{\beta\left(t-l\right)+ u\right\} -  \sin\left\{\beta\left(t-\frac{\tau_{j}}{2}-l\right)+ u\right\}\right]^2.
                       \end{align*}
Using sum to product formula $\sin(a) - \sin(b) = 2\cos\left(\frac{a+b}{2}\right) \sin\left(\frac{a-b}{2}\right)$, we get 
\begin{align*}
    \nu_{j,u}  & = \frac{4\alpha^2}{\tau_{j}^{2}}\frac{1}{M_{j}}\sin^2\left(\frac{\beta \tau_{j}}{4}\right)\sum_{t=\tau_{j}}^{T}\left[\sum_{l=0}^{\frac{\tau_{j}}{2}-1} \cos\left\{\beta\left(t-l-\frac{\tau_{j}}{4}\right)+ u\right\} \right]^2 .
\end{align*}

With sum of cosines formula 
\begin{align*}
\displaystyle \sum_{n=0}^{N-1} \cos \left(a + nd\right)=\left\{
        \begin{array}{lcl}
            N \cos(a) &\text{if}  &\sin\left(\frac{1}{2} d\right) = 0 \\
           R \cos \left\{a + (N-1)\frac{d}{2}\right\} & &\text{otherwise} ,  
        \end{array} 
    \right.   
\end{align*}
where $R = \frac{\sin\left(\frac{1}{2}Nd\right)}{\sin\left(\frac{1}{2}d\right)}$, we obtain
\begin{align*}
    \nu_{j,u}  & =  \frac{4\alpha^2}{\tau_{j}^{2}} \frac{1}{M_{j}} \frac{\sin^4\left(\frac{\beta \tau_{j}}{4}\right)}{\sin^2\left(\frac{\beta}{2}\right)} \sum_{t=\tau_{j}}^{T} \cos^2\left\{\beta\left(t-\frac{\tau_{j}}{2}\right)+\frac{\beta}{2}+ u\right\}.
\end{align*}
Applying power reduction formula $2\cos^2(\theta) = 1+\cos(2\theta)$ to the last equality we derived above, we attain
\begin{align*}
    \nu_{j,u}  & =  \frac{4\alpha^2}{\tau_{j}^{2}} \frac{1}{M_{j}} \frac{\sin^4\left(\frac{\beta \tau_{j}}{4}\right)}{\sin^2\left(\frac{\beta}{2}\right)} \sum_{t=\tau_{j}}^{T} \frac{1+\cos\left\{\beta\left(2t-\tau_{j}\right)+\beta+2 u\right\}}{2}.
\end{align*}
Using power reduction formula $2\sin^2(\theta) = 1-\cos(2\theta)$ to $\sin^4\left(\frac{\beta \tau_{j}}{4}\right)$ and $\sin^2\left(\frac{\beta}{2}\right)$, we get
\begin{align*}
    \nu_{j,u}  & =  \frac{\alpha^2}{\tau_{j}^{2}} \frac{1}{M_{j}} \frac{\left\{1-\cos\left(\frac{\beta \tau_{j}}{2}\right)\right\}^2}{1-\cos\left(\beta\right)} \sum_{t=\tau_{j}}^{T} 1+\cos\left\{\beta\left(2t-\tau_{j}\right)+\beta+2 u\right\}.
\end{align*}
Reusing sum of cosines formula, we can simplify the last equality to obtain
\begin{align*}
    \nu_{j,u}  & =  \frac{\alpha^2}{\tau_{j}^{2}} \frac{\left\{1-\cos\left(\frac{\beta \tau_{j}}{2}\right)\right\}^2}{1-\cos\left(\beta\right)} + \frac{\alpha^2}{\tau_{j}^{2}M_{j}} \frac{\left\{1-\cos\left(\frac{\beta \tau_{j}}{2}\right)\right\}^2}{1-\cos\left(\beta\right)}\frac{\sin\left\{\beta\left(T-\tau_{j}+1\right)\right\}}{\sin\left(\beta\right)}\cos\left\{\beta\left(T+1\right)+2 u\right\}.
\end{align*}
Combining this equality on $\nu_{j,u}$ and the fact that 
\begin{align*}
    \mathbb{E} \left[\cos\left\{\beta(T+1)+2 U\right\}\right] = \frac{1}{2\pi}  \int_0^{2\pi}  \cos\left\{\beta\left(T+1\right)+2u\right\} du = 0,
\end{align*}
where $U\sim\mathcal{U}(0,2\pi)$, we deduce that 
%
\begin{equation*}
    \mathbb{E}\left[\hat{\nu}_j \right] = \frac{\alpha^2}{\tau_{j}^{2}} \frac{\left\{1-\cos\left(\frac{\beta \tau_{j}}{2}\right)\right\}^2}{1-\cos\left(\beta\right)}.
\end{equation*}
%
%
In addition, by Markov's inequality, we have, for any $C > 0$,
\begin{equation*}
\begin{aligned}
        \Pr\left[T \,\big|\,\hat{\nu}_j\left(\alpha, \beta\right) - \mathbb{E}\left\{\hat{\nu}_j\left(\alpha, \beta\right)\right\}\!\big| \geq C \right] &\leq \frac{T^2}{C^2} \mathbb{E}\left(\left[\hat{\nu}_j\left(\alpha, \beta\right) - \mathbb{E}\left\{\hat{\nu}_j\left(\alpha, \beta\right) \right\}\right]^2\right)\\
        &\leq \frac{16 T^2 \alpha^4}{C^2 \tau_{j}^4 M_j^2 \sin^2(\beta) \left\{1 - \cos(\beta)\right\}^2},
\end{aligned}
\end{equation*}
where the second inequality can be deduced form $$\mathbb{E}\left(\left[\hat{\nu}_j\left(\alpha, \beta\right) - \mathbb{E}\left\{\hat{\nu}_j\left(\alpha, \beta\right) \right\}\right]^2\right) = \mathbb{E}\{a^2\cos^2(b+2U)\} = \frac{a^2}{2},$$ where $a\vcentcolon=\frac{\alpha^2}{\tau_{j}^{2}M_{j}} \frac{\left\{1-\cos\left(\frac{\beta \tau_{j}}{2}\right)\right\}^2}{1-\cos\left(\beta\right)}\frac{\sin\left\{\beta\left(T-\tau_{j}+1\right)\right\}}{\sin\left(\beta\right)}$ and $b\vcentcolon=\beta(T+1)$. 
Finally, we have 
\begin{equation*}
\begin{aligned}
        \Pr\left[T \,\big|\,\hat{\nu}_j\left(\alpha, \beta\right) - \mathbb{E}\left\{\hat{\nu}_j\left(\alpha, \beta\right)\right\}\!\big| \geq C \right] 
        &\leq \frac{16 T^2 \alpha^4}{C^2 \tau_{j}^4 M_j^2 \sin^2(\beta) \left\{1 - \cos(\beta)\right\}^2}\\
        &\leq \frac{16 \alpha^4}{C^2 \sin^2(\beta) \left\{1 - \cos(\beta)\right\}^2} \leq \frac{D}{C^2},
\end{aligned}
\end{equation*}

where we used  $\tau_{j}^2 M_{j} > T$ in the first inequality and where $D \vcentcolon= 16 \alpha^4/\left[\sin(\beta) \left\{[1 - \cos(\beta)\right\}\right]^2 < \infty$. Thus, we obtain
\begin{equation*}
    \hat{\nu}_j\left(\alpha, \beta\right) = \mathbb{E}\left\{\hat{\nu}_j\left(\alpha, \beta\right)\right\} + \mathcal{O}_p\left(T^{-1}\right),
\end{equation*}
which concludes the proof.
\begin{flushright}
$\square$
\end{flushright}


\section{Proof of Lemma \ref{lemma:ident:sinus}}
\label{appendix:proof:lemma:ident:sinus}


We define
\begin{align*}
    \bm{\nu}^*(\alpha, \beta) &:= \left[\nu_1(\alpha, \beta) \;,\; \nu_2(\alpha, \beta) \right]^T = \left[\frac{\alpha^2 \{(1-\cos(\beta )\}}{4} \;,\; \frac{\alpha^2 \{1-\cos(2\beta)\}^2}{16 \{1-\cos(\beta)\}} \right]^T ,
\end{align*}

where $ \left[\alpha \;,\; \beta \right]^T \in \bm{\Theta} := \real_+ \times \left(0, \pi\right]$. Fix $\left[\alpha_i \;,\; \beta_i \right]^T \in \bm{\Theta}, \; i = 1,2$ such that $\bm{\nu}^*(\bm{\theta}_1) = \bm{\nu}^*(\bm{\theta}_2)$. Therefore, we have the following system of equations
\begin{equation}
      \label{eq:inter}
        \begin{aligned}
       \frac{\alpha_1^2 \{1-\cos(\beta_1)\}}{4}  &= \frac{\alpha_2^2 \{1-\cos(\beta_2)\}}{4}
      \\
      \frac{\alpha_1^2 \{1-\cos(2\beta_1)\}^2}{16 \{1-\cos(\beta_1)\}} &= \frac{\alpha_2^2 \{1-\cos(2\beta_2)\}^2}{16 \{1-\cos(\beta_2)\}}.
        \end{aligned}
\end{equation}
The first equation of \eqref{eq:inter} implies that
\begin{equation}
      \frac{\alpha_2^2}{\alpha_1^2}  = \frac{1-\cos(\beta_1)}{1-\cos(\beta_2)},
      \label{eq:inter1}
\end{equation}
and the second one that
\begin{equation}
        \frac{\alpha_2^2}{\alpha_1^2} \frac{1-\cos(\beta_1)}{1-\cos(\beta_2)} = \frac{\left\{1-\cos(2\beta_1)\right\}^2}{\left\{1-\cos(2\beta_2)\right\}^2}.
      \label{eq:inter2}
\end{equation}
Using \eqref{eq:inter1} in \eqref{eq:inter2}, we obtain
%



%
\begin{equation}
\left\{\frac{1-\cos(2\beta_2)}{1-\cos(\beta_2)}\right\}^2 = \left\{\frac{1-\cos(2\beta_1)}{1-\cos(\beta_1)}\right\}^2.
    \label{eq:inter3}
\end{equation}
Setting $f(x)=4\{1+\cos(x)\}^2$ and since $\beta \in (0,\pi]$, \eqref{eq:inter3} is equivalent to $f(\beta_2) = f(\beta_1).$
%
%
Since $f(x)$ is injective on $[0,\pi]$, the last equality implies that $\beta_1 = \beta_2$ and using \eqref{eq:inter1}, this also implies that $\alpha_1 = \alpha_2$. Therefore, this proves that $\bm{\nu}^*(\bm{\theta})$ is injective and consequently $\bm{\nu}(\bm{\theta})$ is also injective for $J \geq 2$.
\begin{flushright}
$\square$
\end{flushright}

\section{Proof of Lemma \ref{lemma:ident:sinus&others}}
\label{appendix:proof:lemma:ident:sinus&others}

We first recall the expression of the WV at scale $j\in\{1,\hdots, J\}$, for each process composing $(X_t)$:
\begin{align*}
 \text{White Noise:} \quad&   \nu_{j}(\sigma^{2}) = \frac{\sigma^{2}}{\tau_j},
  \\
   \text{Quantization Noise:} \quad& 
   \nu_{j}(Q^{2}) = \frac{6 Q^{2}}{\tau_j^{2}}, 
  \\
   \text{AR1:} \quad& \nu_{j}(\phi, \varsigma^{2}) =\frac{\varsigma^{2}\left\{\left(\phi^{2}-1\right) \tau_j+2 \phi\left(\phi^{\tau_j}-4 \phi^{\nicefrac{\tau_j}{2}}+3\right)\right\}}{(\phi-1)^{3}(\phi+1) \tau_j^{2}}, 
  \\
   \text{Drift:} \quad& 
   \nu_{j}(\omega^{2})  = \frac{\tau_j^{2} \omega^{2}}{16},
  \\
   \text{Random Walk:} \quad&    \nu_{j}(\gamma^{2}) = \frac{\left(\tau_j^{2}+2\right) \gamma^{2}}{12 \tau_j},
   \\
   \text{Sinusoidal Noise:} \quad&  {{\nu}}_{j}\left(\alpha, \beta\right) = \frac{\alpha^{2}\left\{1-\cos \left(\frac{\beta \tau_{j}}{2}\right)\right\}^{2}}{\tau_{j}^{2}\left\{1-\cos \left(\beta\right)\right\}}, 
\end{align*}
where $\sigma^{2}, Q^{2}, \varsigma^{2}, \omega^{2}, \gamma^{2}, \alpha \in \real_+$, $|\phi|\in(0,1)$ and $\beta\in(0,\pi]$. Set $\bm\theta \vcentcolon= \left[\sigma^{2}, Q^{2}, \phi, \varsigma^{2}, \omega^{2}, \gamma^{2}, \alpha, \beta\right]^T$ and $\bT\subset\real_+^8$ isomorphic to $\real_+^6\times\{(-1,0)\cup(0,1)\}\times(0,\pi]$, the WV at scale $j=1,\dots,J$ of the composite process $(X_t)$ is given by
\begin{align}\label{eq:nu:theta}
    \nu_j(\bm\theta) = \nu_{j}(\sigma^{2}) +\nu_{j}(Q^{2}) + \nu_{j}(\phi, \varsigma^{2}) + \nu_{j}(\omega^{2}) + \nu_{j}(\gamma^{2}) +  \nu_{j}(\alpha,\beta).
\end{align}

The function $\nu_j(\bm\theta)$ can naturally be extended to $\nu_\tau(\bm\theta)$ defined by replacing  $\tau_j$'s entering \eqref{eq:nu:theta} by a variable $\tau\in\real$ (even $\tau\in\mathbb{C}$). The latter function is therefore obtained as the summation of the following functions:
 \begin{equation}
 \begin{split}\label{eq:nu:tau:single:processes}
\nu_{\tau}(\sigma^{2}) &= \frac{\sigma^{2}}{\tau}, 
\\
\nu_{\tau}(Q^{2}) &= \frac{6 Q^{2}}{\tau^{2}}, 
\\
\nu_{\tau}(\phi, \varsigma^{2}) &=\frac{\varsigma^{2}\left\{\left(\phi^{2}-1\right) \tau+2 \phi\left(\phi^{\tau}-4 \phi^{\nicefrac{\tau}{2}}+3\right)\right\}}{(\phi-1)^{3}(\phi+1) \tau^{2}}, 
\\
\nu_\tau(\omega^{2})  &= \frac{\tau_j^{2} \omega^{2}}{16}, 
\\
\nu_{\tau}(\gamma^{2}) &= \frac{\left(\tau^{2}+2\right) \gamma^{2}}{12 \tau},
\\
 {{\nu}}_{\tau}\left(\alpha, \beta\right) &= \frac{\alpha^{2}\left\{1-\cos \left(\frac{\beta \tau_{}}{2}\right)\right\}^{2}}{\tau_{}^{2}\left\{1-\cos \left(\beta\right)\right\}}.
\end{split}
\end{equation}
Given a $\bt\in\bT$, we can consider $\nu_\tau(\bm\theta)$ as functions of $\tau\in\real$ and we will write $\nu_{\bt}(\tau)$ in place of $\nu_{\tau}(\bt)$ in the sequel of the proof.

%
%
Recalling that $\Lambda \vcentcolon=\left[2,2^J\right]\cap\mathbb{Q}$, suppose that for some $\bt_1,\bt_2\in\bT\subset\real^8$, we have that for all $ \tau\in \Lambda$
\begin{equation}\label{eq:ident:mix}
   \nu_{\bt_1}(\tau)=\nu_{\bt_2}(\tau).
\end{equation}

Using \eqref{eq:ident:mix}, we clearly have that $\tau^2\nu_{\bt_1}(\tau)=\tau^2\nu_{\bt_2}(\tau)$, for all $\tau\in\Lambda$. 
We consider the following facts:
\vspace{-0.3cm}
\begin{itemize}
    \item[(i)]  By continuity and the fact that $\Lambda$ is a dense subset of $\left[2,2^J\right]\subset\real$, we have $\forall \tau\in\left[2,2^J\right]$, $\tau^2\nu_{\bt_1}(\tau)=\tau^2\nu_{\bt_2}(\tau)$.
    \item[(ii)] $\forall\bt\in\bT$, $\tau^2\nu_{\bt}(\tau)$ is an analytic function of $\tau\in\real$ (even $\tau\in\mathbb{C}$) since its only non-polynomial expressions in $\tau$ are $\cos(\nicefrac{\beta\tau}{2})$ in the wavelet variance of the sinusoidal process, and $\phi^\tau-4\phi^{\nicefrac{\tau}{2}}$ in the wavelet variance of the AR1 process which are both analytic functions of $\tau\in\real$. Hence, by (i) and the unicity of the analytical continuation, we have that $\forall\tau\in\real$, $\tau^2\nu_{\bt_1}(\tau)=\tau^2\nu_{\bt_2}(\tau)$. 
    \item[(iii)] By (ii), $\tau^2\nu_{\bt_1}(\tau)$ and $\tau^2\nu_{\bt_2}(\tau)$ have the same Taylor expansion for all $\tau\in\real$ and thus the same coefficients.
\end{itemize}
The last step of the proof is to show that (iii) implies that $\bt_1=\bt_2$. From there, we can directly deduce that $\nu_{\bt_1}(\tau)=\nu_{\bt_2}(\tau)$ implies $\bt_1=\bt_2$, which will allow us to complete the proof.





Now, let us compute the Taylor series associated to $\tau^2\nu_{\bt}(\tau)$, where $\bt=\left[\sigma^{2}, Q^{2}, \phi, \varsigma^{2}, \omega^{2}, \gamma^{2}, \alpha, \beta\right]^T$. We start by computing the Taylor series associated to
%
\begin{align*}
    \tau^2 \nu_{\tau}(\alpha,\beta) =
       \displaystyle\frac{\alpha^2 \{1-\cos(\nicefrac{\beta \tau}{2})\}^2}{1-\cos(\beta)}=\frac{\alpha^2}{1-\cos(\beta)}\{1-\cos(\nicefrac{\beta \tau}{2})\}^2.
\end{align*}
First, we have 

\begin{equation*}
\displaystyle1-\cos \left(\nicefrac{\beta\tau}{2}\right)=-\sum_{n=1}^{\infty} \frac{(-1)^{n}}{(2 n) !}\left(\frac{\beta}{2}\right)^{2 n} \tau^{2 n},
\end{equation*}
which gives
\begin{equation*}
\displaystyle\{1-\cos \left(\nicefrac{\beta\tau}{2}\right)\}^2=\sum_{n=2}^{\infty}\left[\sum_{m=1}^{n} \frac{1}{(2 m)!\{2(n-m)\}!}\right](-1)^{n}\left(\frac{\beta}{2}\right)^{2 n}\tau^{2 n}. 
\end{equation*}
Therefore, we have
\begin{equation}\label{eq:taylor:sin}
    \tau^2 \nu_{\tau}(\alpha,\beta)= \sum_{n=2}^{\infty}a_n\tau^{2 n},
\end{equation}
where for all $n\geq 2$ we have 
\begin{equation}\label{eq:coeff:sin}
    a_n = \left[\sum_{m=1}^{n} \frac{1}{(2 m)!\{2(n-m)\}!}\right]\frac{\alpha^2}{1-\cos(\beta)}(-1)^{n}\left(\frac{\beta}{2}\right)^{2 n}.
\end{equation}
For the AR1 process, we first need to express the Taylor series corresponding to $\phi^\tau-4\phi^{\nicefrac{\tau}{2}}$. Since $\phi\in(-1,0)\cup(0,1)$, we have 
\begin{equation*}
    \phi^\tau =  \left\{
    \begin{array}{ll}
        e^{\tau\log(\phi)} & \mbox{if } \phi \in (0,1) \\
        
        e^{\tau \{i\pi+\tau\log(-\phi)\}} & \mbox{if } \phi \in (-1,0),
      \end{array}
\right.
\end{equation*}
where $i\vcentcolon=\sqrt{-1}$ and using the Taylor expansion of the exponential function, we have
\begin{equation*}
    \phi^\tau =  \left\{
    \begin{array}{ll}
        \displaystyle\sum_{n=0}^{\infty} \frac{\ln (\phi)^{n}}{n !}\tau^{n} & \mbox{if } \phi \in (0,1) \\
        
        \displaystyle\sum_{n=0}^{\infty} \frac{\{i\pi+\ln (-\phi)\}^{n}}{n !}\tau^{n} & \mbox{if } \phi \in (-1,0).
      \end{array}
\right.
\end{equation*}
Hence, we can compute the following 
\begin{equation*}
        \tau^2\nu_{\tau}\left(\phi, \varsigma^{2}\right) = \frac{\varsigma^{2}\left\{\left(\phi^{2}-1\right) \tau+2 \phi\left(\phi^{\tau}-4 \phi^{\nicefrac{\tau}{2}}+3\right)\right\}}{(\phi-1)^{3}(\phi+1)} = \displaystyle\sum_{n=1}^{\infty} b_n\tau^{n},
\end{equation*}
where
\begin{equation}\label{eq:a1:AR1:process}
b_1    = \left\{
    \begin{array}{ll}
        \displaystyle\frac{\varsigma^{2}\left\{\phi^{2}-2 \phi \ln (\phi)-1\right\}}{(\phi-1)^{3}(\phi+1)} & \mbox{if } \phi \in (0,1) \\
        
        \displaystyle\frac{\varsigma^{2}\left[\phi^{2}-2 \phi \{i\pi+\ln (-\phi)\}-1\right]}{(\phi-1)^{3}(\phi+1)}  & \mbox{if } \phi \in (-1,0),
    \end{array}
\right.
\end{equation}
and for $n\geq 2$
\begin{equation}\label{eq:an:AR1:process}
b_n   = \left\{
    \begin{array}{ll}
        \displaystyle \frac{\varsigma^{2}\left(1-2^{2-n}\right) 2 \phi}{(\phi-1)^{3}(\phi+1)} \frac{\ln (\phi)^{n}}{n !} & \mbox{if } \phi \in (0,1) \\
        
        \displaystyle \frac{\varsigma^{2}\left(1-2^{2-n}\right) 2 \phi}{(\phi-1)^{3}(\phi+1)} \frac{\{i\pi+\ln (-\phi)\}^{n}}{n !} & \mbox{if } \phi \in (-1,0).
    \end{array}
\right.
\end{equation}

Using equalities \eqref{eq:nu:tau:single:processes}, \eqref{eq:taylor:sin}, \eqref{eq:coeff:sin}, \eqref{eq:a1:AR1:process} and \eqref{eq:an:AR1:process}, for any $\bt=\left[\sigma^{2}, Q^{2}, \phi, \varsigma^{2}, \omega^{2}, \gamma^{2}, \alpha, \beta\right]^T\in\bT$ we have that
\begin{equation}\label{eq:coeff:wv:total}
    \tau^2\nu_{\bt}(\tau) = \displaystyle\sum_{n=0}^{\infty} c_n\tau^{n},
\end{equation}
where
\begin{equation*}
        c_0 = 6 Q^2 , \hspace{0.4cm}  c_1 = \displaystyle \sigma^2 + \frac{\gamma^2}{12} + b_1,  \hspace{0.4cm} c_2 = 0,  \hspace{0.4cm}  c_3 = \displaystyle\frac{\gamma^2}{6} + b_3,
\end{equation*}
and 
\begin{equation*}
      c_4 =  \displaystyle\frac{\omega^2}{16} + a_4+ b_4, \hspace{0.4cm} 
      \\
      c_n = \left\{
    \begin{array}{ll}
             b_n & \mbox{if } n > 4 \mbox{ and } n\in 2\mathbb{N}+1 \\
        
         a_n +b_n & \mbox{if } n > 4 \mbox{ and } n\in 2\mathbb{N}.
        
    \end{array} 
    \right.
\end{equation*}

Now, we consider $\bt_1=\left[\sigma^2_1,\gamma^2_1,\omega^2_1,Q^2_1,\phi_1,\varsigma^2_1,\alpha_1,\beta_1\right]^T$ and $\bt_2=\left[\sigma^2_2,\gamma^2_2,\omega^2_2,Q^2_2,\phi_2,\varsigma^2_2,\alpha_2,\beta_2\right]^T\in\bT$ and their associated Taylor expansion \eqref{eq:coeff:wv:total} denoted as follows
\begin{equation*}
     \tau^2\nu_{\bt_j}(\tau) = \displaystyle\sum_{n=0}^{\infty} c^{(j)}_n\tau^{n},
\end{equation*}
where $j=1,2$. Since \eqref{eq:ident:mix} is satisfied, it implies that for all $n\in\N$ we have $c^{(1)}_n=c^{(2)}_n$. Given the definition of these coefficients, we are going to consider four categories of indices.

\begin{itemize}
    
    \item[I.] \underline{$n=0$}: We clearly have $Q^2_1=Q^2_2$ from $c^{(1)}_0=c^{(2)}_0$.

    \item[II.] \underline{$n > 4$ and $n\in 2\mathbb{N}+1$}: For these indices, $c^{(1)}_n=c^{(2)}_n$ is equivalent to $b_n^{(1)}=b_n^{(2)}$ which are associated uniquely with the AR1 process. From \eqref{eq:an:AR1:process}, three cases need to be considered:
    \begin{enumerate}
        \item \underline{$\phi_1\phi_2 > 0$ and $\phi_1 > 0$}: In this case, $b_n^{(1)}=b_n^{(2)}$ implies 
        \begin{equation*}
          0 < \displaystyle \frac{\varsigma_1^{2} \phi_1}{\varsigma_2^{2} \phi_2}\frac{(\phi_2-1)^{3}(\phi_2+1)}{(\phi_1-1)^{3}(\phi_1+1)} 
            =
            \displaystyle \left\{\frac{\ln (\phi_2)}{\ln (\phi_1)}\right\}^n. 
         \end{equation*}
         Since this equality is supposed to be true for all $n > 4$ with $n\in 2\mathbb{N}+1$, we have
         \begin{equation*}
          \displaystyle \frac{\ln (\phi_2)}{\ln (\phi_1)} = 1 \Rightarrow 
             \phi_1 = \phi_2 
             \Rightarrow 
             \varsigma_1^{2} =\varsigma_2^{2}.
        \end{equation*}

     \item \underline{$\phi_1\phi_2 > 0$ and $\phi_1 < 0$}: Similarly, $b_n^{(1)}=b_n^{(2)}$ for all $n > 4$ with $n\in 2\mathbb{N}+1$ implies
        \begin{equation*}
          0 < \displaystyle \frac{\varsigma_1^{2} \phi_1}{\varsigma_2^{2} \phi_2}\frac{(\phi_2-1)^{3}(\phi_2+1)}{(\phi_1-1)^{3}(\phi_1+1)} 
            =
            \displaystyle \left\{\frac{i\pi + \ln (\phi_2)}{i\pi + \ln (\phi_1)}\right\}^n,
        \end{equation*}
         for all these $n$ and thus gives 
         \begin{equation*}
            \displaystyle \frac{i\pi + \ln (\phi_2)}{i\pi + \ln (\phi_1)} = 1 \Rightarrow 
             \phi_1 = \phi_2 
             \Rightarrow 
             \varsigma_1^{2} =\varsigma_2^{2}.
         \end{equation*}
        \item \underline{$\phi_1\phi_2 < 0$}: We can suppose without loss of generality that $\phi_1 > 0$. Here, $b_n^{(1)}=b_n^{(2)}$ for all $n > 4$ with $n\in 2\mathbb{N}+1$ leads to a contradiction. Indeed, since $\{i\pi + \ln(-\phi_2)\}^2\in\complex\backslash\real$, it implies that $b_n^{(2)}\in\complex\backslash\real$ for an infinite number of $n > 4$  with $n\in 2\mathbb{N}+1$. However, $b_n^{(1)}\in\real$ for all $n\in\N$ which implies that $ b_n^{(1)}\neq b_n^{(2)}$ for an infinite number of $n > 4$  with $n\in 2\mathbb{N}+1$ which is contradictory.  
    \end{enumerate}

\item[III.] \underline{$n > 4$ and $n\in 2\mathbb{N}$}: Here, $c^{(1)}_n=c^{(2)}_n$ is equivalent to $a_n^{(1)} + b_n^{(1)} = a_n^{(2)} + b_n^{(2)}$. However, from the previous case, this is equivalent to $a_n^{(1)} = a_n^{(2)}$ which are uniquely associated with the sinusoidal process. From Lemma \ref{lemma:ident:sinus}, we have $[\alpha_1,\beta_1]^T=[\alpha_2,\beta_2]^T$.

\item[IV.] \underline{$n=1,3,4$}: From the two previous cases, we have that $c^{(1)}_3=c^{(2)}_3$ and $c^{(1)}_4=c^{(2)}_4$ imply $\gamma_1^2=\gamma_2^2$ and $\omega_1^2=\omega_2^2$. From there, $c^{(1)}_1=c^{(2)}_1$ implies $\sigma_1^2=\sigma_2^2$.
\end{itemize}

This concludes the proof.
\begin{flushright}
$\square$
\end{flushright}

\section{Information on Simulation Settings in Section~\ref{section_simu} and Section~\ref{section_application}}
\label{appendix:simu_param_values}

\begin{table}[ht]
\centering
\begin{tabular}{c|l}
Parameter & Value   \\ \hline
      $\phi$   &  $0.975$  \\ 
        $\varsigma^2$ & $0.03\phantom{00}$    \\ 
         $\sigma^2$ & 1\phantom{000.0}   \\
         $\gamma^2$ & $0.0004$\\
         $\alpha$ & $0.85\phantom{00}$\\
         $\beta$ & $0.35\phantom{00}$
\end{tabular}
    \caption{Parameters for Simulation Study 1}
    \label{tab:simu_1}
\end{table}

\begin{table}[ht]
    \centering
\begin{tabular}[t]{c|l}
Parameter & Value   \\ \hline
      $\phi_1$   &  $0.999995$  \\ 
        $\varsigma^2_1$ & $3\cdot 10^{-11}$    \\ 
         $\phi_2$ & $0.1107083$   \\
         $\varsigma^2_2$ & $5.278666\cdot 10^{-4}$\\
         $\alpha_1$ & $0.025$\\
         $\beta_1$ & $0.056$ \\
         $\alpha_2$ & $0.0015$\\
         $\beta_2$ & $8\cdot 10^{-5}$ 
         
\end{tabular}
    \caption{Parameters for Simulation Study 2}
    \label{tab:simu_2}
\end{table}

\begin{table}[ht]
    \centering
  \begin{tabular}[t]{c|l}
 Parameter & Value   \\ \hline
      $\phi$   &  $0.9997083$  \\ 
        $\varsigma^2$ & $9\cdot 10^{-9}$  \\ 
         $\sigma^2$ & $8\cdot 10^{-4}$   \\
         $\gamma^2$ & $3\cdot 10^{-11}$\\
         $\alpha$ & $0.025$\\
         $\beta$ & $0.056$
\end{tabular}

    \caption{Parameters for Simulation Study 3}
    \label{tab:simu_3}
\end{table}

\begin{table}[ht]
    \centering
    \begin{tabular}[t]{c|l}

 Parameter & Value   \\ \hline
 \rule{0pt}{3ex}    
      $\phi$   &  $1.851173\cdot 10^{-01}$  \\ 
        $\varsigma^2$ & $3.559081\cdot 10^{-02}$  \\ 
        $\gamma^2$ & $8.692479\cdot 10^{-10}$\\
         $\alpha_1$ & $3.235864\cdot 10^{-01}$\\
         $\beta_1$ & $1.199147 $ \\
         $\alpha_2$ & $1.359012\cdot 10^{-01}$\\
         $\beta_2$ & $1.357501\cdot 10^{-01}$ \\
\end{tabular}
    \caption{Parameters for the Emulation Study}
    \label{tab:emulation_study}
\end{table}


\mycomment{For Simulation~1 which compare the GMWM to the MLE for different sample sizes, we define the parameters of the model as such: $\phi_{\mathrm{AR1}} = 0.975$, $\varsigma^2_{\mathrm{AR1}} = 0.03$, $\sigma^2_{\mathrm{WN}} = 1$, $\gamma^2_{\mathrm{RW}} =  0.0004$, $\alpha_{\mathrm{SIN}} = 0.85$ and $\beta_{\mathrm{SIN}} = 0.35$.

Using the superscript ${}^{(i)}$ to denote the parameters of the $i^{th}$ AR1 or sinusoidal process in the model, the parameter values in the Simulation 2 setting are as follows: $\phi_{\mathrm{AR1}}^{(1)} = 0.999995$, $\varsigma^2_{\mathrm{AR1}}^{(1)} = 3\mathrm{e}{-11}$, $\phi_{\mathrm{AR1}}^{(2)} = 0.1107083$, $\varsigma^2_{\mathrm{AR1}}^{(2)} = 5.278666\mathrm{e}{-4}$, $\alpha_{\mathrm{SIN}}^{(1)} =0.025$, $\beta_{\mathrm{SIN}}^{(1)} = 0.056$, $\alpha_{\mathrm{SIN}}^{(2)} = 0.0015$ and $\beta_{\mathrm{SIN}}^{(2)} = 8\mathrm{e}{-5}$. 

For Simulation 3, we fix the parameters of the model as such: $\phi_{\mathrm{AR1}} = 0.9997083$, $\varsigma^2_{\mathrm{AR1}} = 9\mathrm{e}{-9}$, $\sigma^2_{\mathrm{WN}} = 8\mathrm{e}{-4}$, $\gamma^2_{\mathrm{RW}} = 3\mathrm{e}{-11}$, $\alpha_{\mathrm{SIN}} = 0.025$ and $\beta_{\mathrm{SIN}} = 0.056$.  }

\mycomment{For the simulation described in Section~\ref{section_application}, we consider a model composed of the combination of a autoregressive process of order 1 with parameter $\phi_{\mathrm{AR1}} = 1.851173\mathrm{e}{-01}$ and $\varsigma^2_{\mathrm{AR1}} =  3.559081\mathrm{e}{-02}$ coupled with a random walk with parameter $\gamma^2_{\mathrm{RW}} = 8.692479\mathrm{e}{-10}$ and two sinusoidal signal with respectively parameter $\alpha_{\mathrm{SIN-1}} = 3.235864\mathrm{e}{-01}$, $\beta_{\mathrm{SIN-1}} = 1.199147\mathrm{e}{+00} $ and  $\alpha_{\mathrm{SIN-2}} = 1.359012\mathrm{e}{-01}$ and $\beta_{\mathrm{SIN-2}} =   1.357501e-01$.}

\begin{figure}[ht]
    \centering
    \includegraphics[scale=.5]{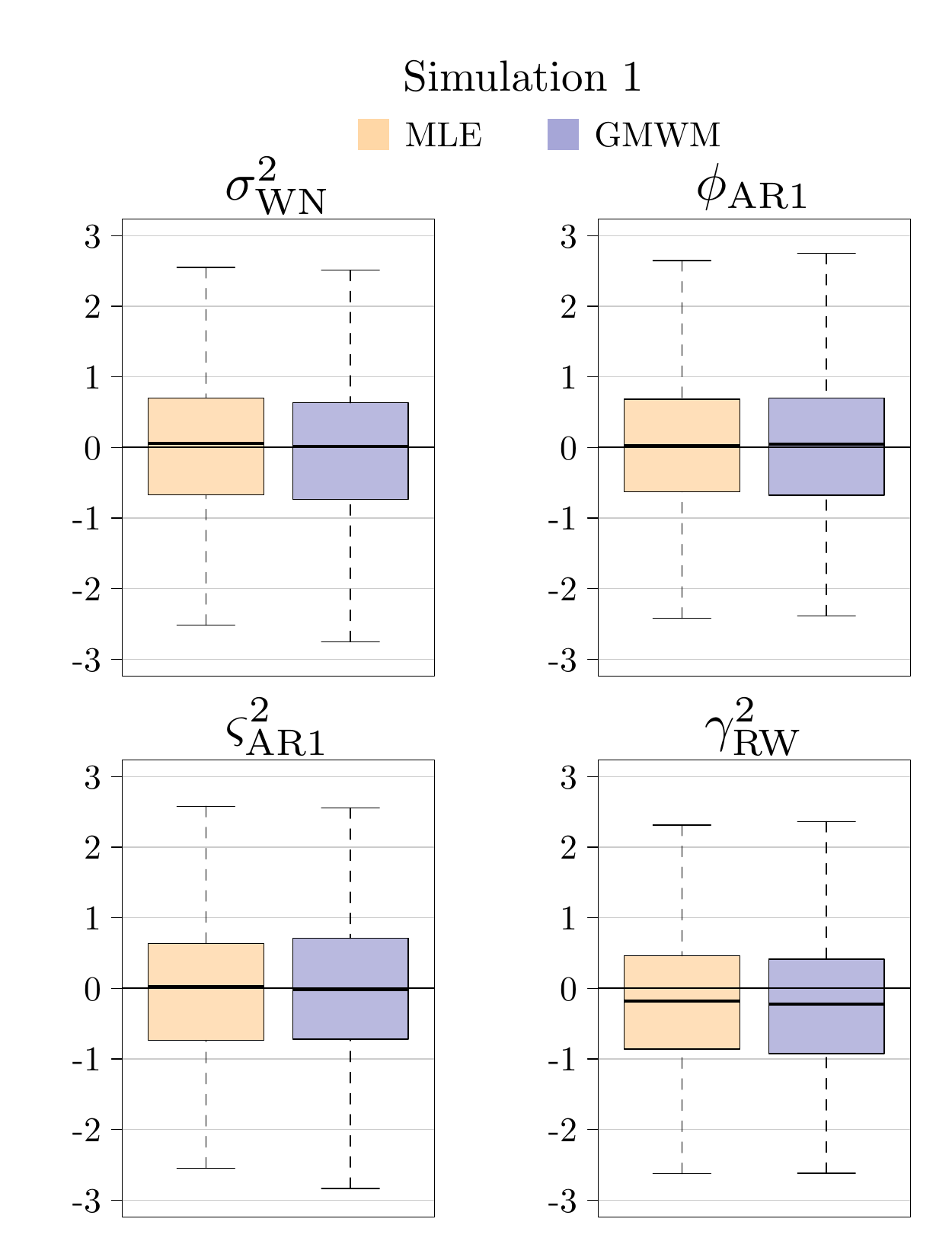}
    \caption{
    Empirical distribution of GMWM and MLE parameter estimates for Simulations 1 for a sample size of $16 \cdot 10^4$. The
true parameter values were subtracted from each boxplot and all distributions were standardized
by their respective empirical variances.}
    \label{fig:boxplot_estimated_param_160000}
\end{figure}

The empirical distribution of the {\it{centered}} and {\it{scaled}} estimated parameters (i.e., the estimated parameter minus its true value divided by its empirical standard deviation) are shown for the largest sample size in Figure~\ref{fig:boxplot_estimated_param_160000}. As it can be observed, the MLE is more efficient than the GMWM as the estimated parameters present a lower variance compared to the estimated parameters obtained with the GMWM, more notably for the parameter of the white noise and the variance parameter of the autoregressive process of order $1$. However, the variability of the GMWM estimator for all parameters is in a comparable magnitude to the variability of the MLE.

\end{document}